\documentclass[final,,twoside]{IEEEtranTCOM}
\normalsize
\ifCLASSINFOpdf
\else
\fi

\usepackage{comment}

\usepackage{amsthm}
\usepackage{amsmath}
\usepackage{amssymb}
\usepackage{graphicx}
\usepackage{esint}
\usepackage{amsfonts}
\usepackage{cite}
\usepackage{balance}
\usepackage{caption}
\usepackage{subcaption}
\usepackage{epstopdf}
\usepackage{color}
\usepackage{tikz}
\usepackage{longtable}

\usepackage{tabularray}
\usepackage{multirow}
\usepackage{suffix}
\usepackage{mathtools}

\makeatletter

\usepackage[scr=esstix,cal=boondox]{mathalfa}

\theoremstyle{plain}
\newtheorem{thm}{\protect\theoremname}

\makeatother

\providecommand{\theoremname}{Theorem}

\theoremstyle{plain}
\newtheorem{lem}{\protect\lemmaname}

\theoremstyle{plain}
\newtheorem{rem}{\protect\remarkname}

\makeatother

\providecommand{\lemmaname}{Lemma}
\providecommand{\remarkname}{Remark}

\DeclarePairedDelimiterX\MeijerM[3]{\lparen}{\rparen}%
{\begin{smallmatrix}#1 \\ #2\end{smallmatrix}\delimsize\vert\,#3}

\newcommand\MeijerG[8][]{%
  G^{\,#2,#3}_{#4,#5}\MeijerM[#1]{#6}{#7}{#8}}

\WithSuffix\newcommand\MeijerG*[7]{%
  G^{\,#1,#2}_{#3,#4}\MeijerM*{#5}{#6}{#7}}

  \DeclarePairedDelimiterX\FoxM[3]{\lparen}{\rparen}%
{\begin{smallmatrix}#1 \\ #2\end{smallmatrix}\delimsize\vert\,#3}

\newcommand\FoxH[8][]{%
  H^{\,#2,#3}_{#4,#5}\FoxM[#1]{#6}{#7}{#8}}

\WithSuffix\newcommand\FoxH*[7]{%
  H^{\,#1,#2}_{#3,#4}\FoxM*{#5}{#6}{#7}}

\begin{document}
\title{On the Reliability and Security of Ambient Backscatter Uplink NOMA Networks}


\author{Athanasios P. Chrysologou,~\IEEEmembership{Graduate Student Member,~IEEE,} Nestor D. Chatzidiamantis,~\IEEEmembership{Member,~IEEE,} Alexandros-Apostolos A. Boulogeorgos,~\IEEEmembership{Senior Member,~IEEE}, and Zhiguo Ding,~\IEEEmembership{Fellow, IEEE}
 \thanks{A. P. Chrysologou and N. D. Chatzidiamantis are with the Department of Electrical and Computer Engineering, Aristotle University of Thessaloniki, 54124 Thessaloniki, Greece (e-mails:
chrysolog@ece.auth.gr, nestoras@auth.gr).} 
\thanks{A.-A. A. Boulogeorgos is with the Department of Electrical and Computer Engineering, University of Western Macedonia, 50100 Kozani, Greece (e-mail:
al.boulogeorgos@ieee.org).} 
\thanks{Z. Ding is with the Department of Electrical Engineering and
Computer Science, Khalifa University, Abu Dhabi, United Arab Emirates,
and also with the Department of Electrical and Electronic Engineering, The University of Manchester, M13 9PL Manchester, U.K. (e-mail:
zhiguo.ding@manchester.ac.uk).
}

\thanks{Part of this work has been presented at IEEE 97th Vehicular Technology Conference (VTC2023-Spring) \cite{chrysologouconf}.}

\thanks{The research was supported by MINOAS project within the H.F.R.I call ''Basic research Financing (Horizontal support of all Sciences)'' under the National Recovery and Resilience Plan ``Greece 2.0'' funded by the European Union - NextGenerationEU (H.F.R.I. Project Number: 15857).}
	
\vspace{-0.5cm}
}

\maketitle	
\begin{abstract} A fundamental objective of the forthcoming sixth-generation wireless networks is to concurrently serve a vast array of devices many of which, such as Internet-of-Things (IoT) sensors, are projected to have low power requirements or even operate in a battery-free manner. To achieve this goal, non-orthogonal multiple access (NOMA) and ambient backscatter communications (AmBC) are regarded as two pivotal and promising technologies. In this work, we present a novel analytical framework for studying the reliability and security of uplink NOMA-based AmBC systems. Specifically, closed-form analytical expressions for both NOMA-users' and IoT backscatter device's (BD's) outage probabilities (OPs) are derived for both cases of perfect and imperfect successive interference cancellation (SIC). In addition, assuming that one NOMA-user transmits an artificial noise in order to enhance system's security, the physical layer security (PLS) of the  system is investigated by extracting analytical expressions for NOMA-users' and BD's intercept probabilities (IPs). To gain insightful understandings, an asymptotic analysis is carried out by focusing on the high signal-to-noise (SNR) regime, which reveals that NOMA-users and BDs face outage floors in the high SNR regime as well as that IPs reach constant values at high SNR. Additionally, practical insights regarding how different system parameters affect these OP floors and IP constant values are  extracted. Numerical results verify the accuracy of othe developed theoretical framework, offer performance comparisons between the presented NOMA-based AmBC system and a conventional orthogonal multiple access-based AmBC system, and reveal the impact of different system parameters on the reliability and security of NOMA-based AmBC networks.

\end{abstract}
\vspace{-0.1cm}

\vspace{-0.4cm}
\section{Introduction}\label{S:Intro}

\IEEEPARstart{A} significant challenge of the upcoming sixth-generation (6G) networks is the imperative need to concurrently accommodate a vast number of devices and sensors with diverse demands, varying quality of experience (QoE) needs, and assorted energy requirements \cite{zhang20196g,tataria20216g}. Projections indicate that in the forthcoming 6G networks, Internet-of-Everything (IoE) devices will reach densities of at least $10^7$ devices/km$^2$ \cite{zhang20196g}, highlighting the pressing demand for extensive connectivity, spectral efficiency, and energy conservation. In this context, considerable research effort has been directed towards evolving conventional multiple access techniques into the so-called next-generation multiple access (NGMA), which promises to address the multifaceted requirements of 6G networks \cite{liu2022evolution,diamantoulakis2022next}. In particular, non-orthogonal
multiple access (NOMA), whose basic principle is to allow plenty of devices to be served simultaneously in the most spectral efficient way by sharing the same frequency, time or even code resources, is anticipated to play a pivotal role in the 6G and in the frameworks of International Mobile Telecommunications for 2030 and beyond (IMT-2030) \cite{liu2022evolution,liu2022developing,recommendation2023framework}. On a parallel avenue, beyond conventional non-energy-constrained communication scenarios, the vision for the next generation of Internet-of-Things (NGIoT) additionally includes supporting energy-constrained devices \cite{ngiot,9261963}. In the meantime, 3rd Generation Partnership Project (3GPP) is currently working towards the realization of the concept of ambient power-enabled Internet-of-Things (IoT) \cite{chen20233gpp}. In pursuit of these objectives, ambient backscatter communications (AmBC) have garnered attention from both academia and industry. In a typical AmBC scenario, the so-called backscatter devices (BDs) re-modulate and reflect pre-existing radio frequency signals (e.g., cellular or WiFi signals) by changing their antenna impedance.

The aforementioned advantages have sparked heightened research interests in combining AmBC with NOMA. In~\cite{zhang2019backscatter}, a downlink NOMA-assisted AmBC system setup consisting of two cellular users and a passive IoT device was presented and investigated, while, in the meantime, for the same system setup, \cite{li2022effective} carried out an effective capacity analysis and demonstrated that increased BD's reflection coefficient leads to increased effective capacity for the BD but decreased effective capacities for the NOMA-users. In contrast to the previous contributions that assumed a constant reflection coefficient for the BDs, in \cite{yang2023novel}, a downlink NOMA system setup capable of supporting AmBC was considered, where the BD can dynamically adjust its reflection coefficient. This approach was proved to significantly improve the outage performance for both NOMA-users and the BD. The authors of \cite{ding2022symbiotic} assumed a symbiotic AmBC framework consisting of one primary and one backscatter system and assessed the coexistence capability of this system configuration by deriving the system's coexistence outage probability. Additionally, \cite{ding2022symbiotic} analyzed the ergodic capacity of the backscatter link. In \cite{le2021joint}, the analysis was extended to the cognitive radio scenario, wherein the operation of a secondary network in NOMA-AmBC systems was enabled while ensuring protection to the operations of the primary network. It was illustrated that, through appropriate parameter optimization, this system model is capable of providing enhanced energy as well as spectrum efficiency. In \cite{chen2021backscatter}, a cooperation scheme for downlink NOMA-AmBC networks was presented, where a near user backscatters the surplus power of its received signal in order to assist the reception of a far user initially unable to recover its intended messages. It was proved that the proposed backscatter cooperation NOMA scheme offers decreased outage probabilities (OPs) as well as enhanced expected rates compared to conventional non cooperative approaches. Moreover, \cite{elsayed2021noma} and \cite{elsayed2022symbiotic} considered an  AmBC communication scenario, where a source multiplexes two different signals for a single receiver under the coexistance of a BD, and investigated system's outage and ergodic capacity (EC) performance, respectively. 
In \cite{ding2021harvesting}, a comparison between backscatter communication-assisted NOMA and wireless power transfer-assisted NOMA systems in terms of users' OPs and ECs was conducted.

In comparison with the aforementioned works which mainly focused on the performance analysis of NOMA-enabled AmBC networks, research on the system performance optimization in this type of networks has been performed, as well. In \cite{khan2021backscatter} and in \cite{xu2020energy} a joint transmit power and reflection coefficient optimization framework and an energy efficient resource allocation problem with QoS guarantee, respectively, were formulated and solved for the same system setup with \cite{zhang2019backscatter,li2022effective,yang2023novel}. In \cite{ding2021application}, an uplink throughput maximization problem was formulated and solved in a communication scenario with one downlink user and multiple uplink BDs. Extending the analysis to multi-user scenarios, \cite{el2023multi} derived the information-theoretic achievable rate region for both the discrete memoryless and the Gaussian channel. Furthermore, an optimization framework for maximizing the system's energy efficiency was formulated and solved. In the same direction, in \cite{asif2022energy}, a low-complexity optimization algorithm was introduced for the maximization of energy efficiency of cooperative downlink NOMA-AmBC networks. Specifically, the proposed scheme optimizes both source's and relay's transmit power as well as the power allocation coefficients while taking into account users' quality of service (QoS), power budgets, and cooperation constraints.


Meanwhile, because of their wireless transmissions, AmBC networks are vulnerable to eavesdropping. It's worth noting that traditional encryption technologies often involve high computational complexity \cite{li2019secure}, making them impractical for deployment on BDs with limited storage and computing capabilities. Consequently, these encryption methods may not provide viable solutions for addressing the security challenges inherent in AmBC systems. Instead, physical layer security (PLS), whose basic principle is to leverage the inherent randomness of dynamic fading channels in order to facilitate secure communication over wireless mediums without relying on conventional encryption \cite{mukherjee2014principles}, has emerged as a promising approach to increase the security of AmBC wireless systems. In this context, \cite{li2021physical} and \cite{li2023physical} focused their analysis on the PLS of AmBC networks. In \cite{li2021physical} the PLS of a cognitive AmBC network was investigated by extracting uses' and BDs' intercept probabilities (IPs). It was rigorously proved that there exists a trade-off between reliability and security in AmBC networks. Furthermore, \cite{li2023physical} delved into the case of AmBC cooperative networks where a dedicated relay aids transmissions to distant users. It was revealed that system parameters impact the secrecy performance and energy efficiency of the system in unique ways, highlighting the necessity for meticulous parameter selection to ensure the seamless operation of AmBC networks. In the meantime, \cite{li2020secrecy} and \cite{li2021hardware} extended their analysis for the NOMA case by investigating the PLS of downlink NOMA-enabled AmBC systems. In particular, \cite{li2020secrecy} provided a secrecy analysis of such systems under I/Q imbalance, while \cite{li2021hardware} derived users' OPs and IPs for such system models under both residual hardware impairments (RHIs) and channel estimation errors (CEEs), illustrated that RHIs pose a more significant impact on both the reliability and security of the system than CEEs, and pointed out an interesting trade-off between system's reliability and security. 


The above discussions reveal that the great majority of the literature predominantly concentrates on downlink NOMA-assisted AmBC system configurations or point-to-point communication setups involving one or multiple BDs. Consequently, the interplay between uplink NOMA and AmBC networks remains largely unexplored. However, considering the crucial significance of the uplink for various emerging 6G applications, such as NGIoT, augmented and virtual reality, connected robotics and autonomous systems, telemedicine, among others, and recognizing that, according to AmBC principles, battery-free BDs are able to utilize existing uplink signals to transmit their own, the investigation of uplink NOMA-enabled AmBC networks is more relevant than ever.

Meanwhile, the effective deployment of NOMA significantly depends on successful successive interference cancellation (SIC) to surpress multi-access interference \cite{7973146}. Prior studies on NOMA-assisted AmBC networks often assume that receivers are capable of accurately decoding strong interfering signals and then filter them out from the received signal before decoding the target signal. Nevertheless, achieving precise perfect SIC (pSIC) in practical systems poses many challenges; thus, it is anticipated that practical applications will most commonly encounter imperfect SIC (ipSIC), resulting in significant performance degradation \cite{8125754,bisen2021performance,chrysologouSIC}. Furthermore, despite security being reported as one of the most critical concerns for the next-generation networks, and despite vast investigations into the PLS of downlink NOMA-empowered AmBC systems, to the best of the authors' knowledge, there is currently no existing work focusing on the PLS of uplink NOMA-based AmBC networks, which consist of uplink users and passive BDs.
 
Recognizing the aforementioned research gaps, in this work, we develop a novel analytical framework for investigating the reliability and the PLS of uplink NOMA systems that are able to support AmBC. Specifically, the contribution of this paper is summarized below:
\begin{itemize}
    \item We consider an uplink NOMA-based AmBC scenario, where NOMA-users intent to transmit their messages to a base station (BS), while in the meantime, IoT battery-free BDs utilizes the undergoing messages to transmit their information to the BS. In order to enhance the security of the proposed system setup against malicious eavesdropping, we assume that one of the NOMA-users emits an artificial noise.

    \item For the presented system setup, closed-form analytical expressions for NOMA-users' as well as BD's OPs are derived under the case of pSIC as well as for the more practical scenario of ipSIC. 

    \item A PLS investigation of the proposed system model is performed by extracting analytical expressions for the IPs of NOMA-users and BDs.

    \item An asymptotic analysis is also carried out. Specifically, it is revealed that both NOMA-users and BD show outage floors in the high signal-to-noise ratio (SNR) regime for both pSIC and ipSIC cases. Furthermore, it is proved that NOMA-users' and BD's IPs also tend to reach constant values at high SNR. Useful remarks regarding how different system parameters affect the appearing floors are also provided. 
    
    \item Simulation results validate the accuracy of the presented analysis, present a performance comparison between the proposed NOMA-enabled AmBC system model and a conventional orthogonal multiple access (OMA) approach, and illustrate the effect of different system parameters on the NOMA-users' and BD's OPs and IPs.   
    
\end{itemize}

The remainder of the paper is organized as follows: Section II introduces the proposed system model. In Section III, users' and BD's outage performance is investigated by deriving analytical expressions for the OPs, while in Section IV, analytical derivations for users' and BD's IPs are extracted in order to analyze network's  PLS. In Section V, simulation results are presented to validate the accuracy of the provided theoretical results and to illustrate the impact of different parameters on system's performance. Finally, Section VI concludes the paper.

\vspace{-0.2cm}
\section{System model}\label{sec:SSM}

\begin{figure} 
	\centering	\includegraphics[keepaspectratio,width=0.7\columnwidth]{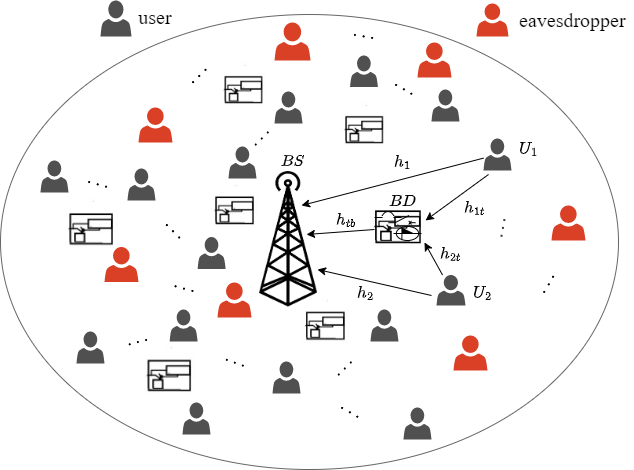}
	\caption{System model.}  
	\label{Fig:sys}	\vspace{-0.45cm}
\end{figure} 

As depicted in Fig. \ref{Fig:sys}, an uplink NOMA-based AmBC system setup is considered, which consists of a BS, $N$ legitimate users, $T$ IoT battery-free BDs and $M$ eavesdroppers (Eves). Legitimate users intent to transmit their messages to the BS exploiting uplink NOMA, while BDs take advantage of the uplink users' transmissions in order to also transmit their message to the BS over their received messages. It is assumed that all nodes are equipped with a single antenna. The channel coefficient from the $i$-th user to the BS is denoted by $h_i$, with $i \in \{1, 2, \cdots,N\}$, the channel between the $i$-th user and the $t$-th BD is denoted by $h_{it}$, $t \in \{1, 2,\cdots, T\}$, the channel between the $i$-th user and the $j$-th eve is denoted by $h_{ij}$, $j \in \{1, 2,\cdots,M\}$, and the channel between the $j$-th eve and the $t$-th BD is denoted by $h_{jt}$. All channels are assumed to be subject to Rayleigh fading, i.e., $h_{i}\!\sim\! CN(0,\lambda_{i})$, $h_{ij}\sim CN(0,\lambda_{ij})$ and $h_{jt}\sim CN(0,\lambda_{jt})$, where $\lambda_{i}$, $\lambda_{ij}$ and $\lambda_{jt}$ denote the variances of the complex normal distributions.   

In order to mitigate the strong inter-user interference of NOMA and the long processing delay of SIC, the users are divided into multiple orthogonal two-user pairs, where NOMA is implemented within each pair and different pairs are served in different time slots \cite{cao2020security,10286123}. Likewise, towards the enhancement of connectivity levels \cite{ding2021harvesting}, each time users of a NOMA-cluster transmit their messages using a unique orthogonal resource block, then an IoT BD is granted access to this orthogonal block. Hereinafter, we focus the analysis on investigating the outage performance of the NOMA-users and the BD of such a NOMA-cluster. For convenience, we denote the legitimate users of such a cluster by $U_1$, $U_2$ and, without loss of generality, we assume that the distance between $U_2$ and the BS is shorter than the distance between $U_1$ and the~BS.    

Following the uplink NOMA principle, $U_{1}$ and $U_{2}$ simultaneously transmit their uplink messages, $x_{1}$, $x_{2}$, respectively, to the BS with $E[|x_1|^2]\!=\!E[|x_2|^2]\!=\!1$, where $E[\cdot]$ denotes expectation. However, in order to improve the security levels of the system, one of the two users is randomly selected to also transmit an artificial noise message $z_t$, with $E[|z_t|^2]\!=\!1$. In the meantime, consider there exists a BD that has been granted access in the resource block of this NOMA-cluster and symbiotically coexists in the network. The BD takes advantage of the undergoing transmissions and backscatters its received message to the BS with its own message $x_t$. Hence, the signal received from the BS can be modeled as
\begin{equation}
	\begin{split}
		y_{b}&= \sqrt{a_1 P} h_{i^{*}} x_{i^{*}} + \sqrt{a_2 P} h_{i^{*}} z_t + \sqrt{P} h_{j^{*}} x_{j^{*}} \\
		&+ \sqrt{\eta a_1 P} h_{{i^{*}}t} h_{tb} x_{i^{*}} x_t + \sqrt{\eta a_2 P} h_{{i^{*}}t} h_{tb} z_t x_t \\ 
		&+ \sqrt{\eta P} h_{j^{*}t} h_{tb} x_{j^{*}} x_t + n_b,
	\end{split} \label{y_bs}
\end{equation}
where $i^{*} \in \{1,2\}$ denotes the user selected to transmit the artificial noise, i.e., the user selected to act as a friendly jammer, $j^{*} \!=\! \{1,2\}\!-\!\{i^{*}\}$, $P$ is the users' transmit power, $\eta$ denotes the BD's reflection coefficient with $0\!\leq\! \eta\! \leq\! 1$ \cite{ding2021application} and $n_b$ is the additive white Gaussian noise (AWGN) coefficient at the BS with zero mean and variance $N_0$. Furthermore, as shown in (\ref{y_bs}), the total power $P$ of user $U_{i^{*}}$ is splitted into two parts, namely $a_1P$ and $a_2P$, used for the transmission of the uplink message $x_{i^{*}}$ and artificial noise $z_t$, respectively; thus, $a_1+a_2=1$. Of note the artificial noise is generated via a pseudorandom sequence, which is known to the BS; thus, the latter is able to cancel the interference from the artificial noise out \cite{cao2020security}. 

Motivated by the fact that in uplink NOMA it is highly important to maintain the distinctness of received signals \cite{7557079} and given the double-fading effect in BD's cascade channel as well as the generally small $\eta$ value \cite{li2022effective}, the decoding order at the BS is considered as $(x_{2},x_{1},x_t)$. As a result, in the ipSIC scenario, the received signal-to-interference-plus-noise ratios (SINRs) at the BS for the decoding of messages $x_2$, $x_1$ and $x_t$, can be respectively obtained as
\begin{equation}
	\begin{split}
	&\gamma_{x_2}\!=\!\frac{ (1 \!-\! \epsilon(1\!-\!a_1)) \rho |h_2|^2}{(1 \!- (1\!-\!\epsilon)(1\!-\!a_1))\rho |h_1|^2\! +\! \eta \rho |h_{tb}|^2 (|h_{2t}|^2\!+\!|h_{1t}|^2)  \! +\! 1}, 
	\end{split} \label{gamma_x2}
\end{equation}
\begin{equation}
\begin{split}
	&\gamma_{x_1}\!=\!\frac{(1 \!-\! (1\!-\!\epsilon)(1\!-\!a_1)) \rho |h_1|^2}{ \eta \rho |h_{tb}|^2 (|h_{2t}|^2\!+\!|h_{1t}|^2) \!+\! (1 \!-\! \epsilon(1\!-\!a_1)) k_2 \rho |h_2|^2  \!+\! 1}, 
 \end{split} \label{gamma_x1}
\end{equation}
\begin{equation}
	\begin{split}
	&\gamma_{x_t}\!=\!  \frac{\eta \rho |h_{tb}|^2 (|h_{2t}|^2\!+\!|h_{1t}|^2)}{ (1 \!-\! (1\!-\!\epsilon)(1\!-\!a_1)) k_1 \rho |h_1|^2 \!+\! (1 \!-\! \epsilon(1\!-\!a_1)) k_2 \rho |h_2|^2   \!+\!   1}\!,
	\end{split}    \label{gamma_xt}
\end{equation}
where $\rho \!=\! \frac{P}{N_0}$ is users' transmit SNR and  $k_1$, $k_2$ denote the impact of the ipSIC related parameters. Note that, for $n \!\in\! \{1,2\}$, it holds $0 \leq\! k_n\! \leq 1$, where the case $k_n\!=\!0$ refers to pSIC, and the case $k_n\!=\!1$ refers to no SIC. Also, $\epsilon$ is a two-state logical variable that becomes equal to zero when $U_1$ is randomly chosen to emit the artificial noise and equal to one when $U_2$ is randomly chosen to emit the artificial noise, i.e.,
\begin{equation}
	\epsilon= \begin{cases}
		0, \underset{i \in \{1,2\}}{\text{rand}}{U_i}=U_1, \text{i.e.,} \ i^{*}=1 
		\\
		1, \underset{i \in \{1,2\}}{\text{rand}}{U_i}=U_2, \text{i.e.,} \ i^{*}=2 
	\end{cases}
\end{equation}

On the other hand, for the Eves, we consider the worst-case scenario, according to which Eves are endowed with sophisticated decoding capabilities and are able to perform parallel interference cancellation (PIC), i.e., every Eve is able to detect each of the messages $\{x_2,x_1,x_t\}$ without being interfered by the other two messages but only from the artificial noise $z_t$ \cite{cao2020security,chen2018physical}. The aforementioned assumption might overestimate Eves' capabilities by providing  an upper bound on their intercept performance; however, this is a practice commonly used for PLS investigation since it allows the design of robust practical applications \cite{cao2020security,chen2018physical}. Building upon the above assumptions, the SINRs at Eves to detect messages $x_2$, $x_1$ and $x_t$ can be respectively expressed~as
\begin{equation}
	\gamma_{2j}=\frac{ (1 - \epsilon(1-a_1)) \rho |h_{2j}|^2}{ a_2 \rho|h_{i^{*}j}|^2  + 1},  \label{g_2j}
\end{equation}
\begin{equation}
	\gamma_{1j}=\frac{ (1 - (1-\epsilon)(1-a_1)) \rho |h_{1j}|^2}{ a_2 \rho|h_{i^{*}j}|^2  + 1},   \label{g_1j}
\end{equation}
\begin{equation}
	\gamma_{tj}=\frac{ \eta \rho |h_{tj}|^2 (|h_{2t}|^2+|h_{1t}|^2)}{ a_2 \rho|h_{i^{*}j}|^2  + 1},    \label{g_tj}
\end{equation}
where $j\! \in\! \{1,...,M\}$ and $i^{*} \!\in\! \{1,2\}$ denotes the randomly selected user that acts as a friendly jammer. 

\vspace{-0.4cm}
\section{Outage Analysis}

Considering that $U_1$, $U_2$ and the BD transmit data at constant data rates, denoted as $R_1$, $R_2$ and $R_t$, respectively, then the OP becomes a significant metric for performance evaluation. This section focuses on deriving precise closed-form expressions for the OPs of all messages within the proposed system model. Additionally, valuable insights regarding the outage performance of $U_1$, $U_2$, and BD in the high SNR regime are presented.

\vspace{-0.2cm}
\subsection{Outage performance of $U_{2}$}

An outage event for $U_2$ occurs when the combined interference from $U_1$ and BD along with the AWGN, prevents the BS from decoding $x_2$. Therefore, the OP of $U_2$ is defined~as
\begin{equation}
	P_{2}^{o} = \Pr \left( \gamma_{x_2} < u_2  \right), \label{OP_2}	
\end{equation}
where $u_2\!=\!2^{R_{2}}\!-\!1$ is the SINR threshold for successful decoding of $x_2$. It is noted that, given the fact that $x_2$ is decoded first at the BS side, ipSIC has no impact on $U_2$'s outage performance. In what follows, we provide a theorem about the OP of $U_{2}$.
\begin{thm}
The OP of $U_2$ can be calculated as
\begin{equation}
	P_{2}^{o}=   1 - \frac{1}{2} \left( I_{1}|_{\epsilon = 0}    + I_{1}|_{\epsilon = 1}  \right)   ,   \label{x2_OP}
\end{equation}
where $I_{1}$ is given at the top of the next page with $A\!=\!1 \!-\! \epsilon(1\!-\!a_1)$, $B\!=\!1 \!-\! (1\!-\!\epsilon)(1\!-\!a_1)$ and $W_{b,c}(\cdot)$ denoting the Whittaker function \cite[(9.220.4)]{grad}. By  $I_{1}|_{\epsilon = 0}$, $I_{1}|_{\epsilon = 1}$ we denote that we set $\epsilon\!=\!0$ or $\epsilon\!=\!1$, respectively, in the expression of $I_1$.   
\begin{figure*} \vspace{-0.2cm}
\begin{equation}
	I_{1}=\begin{cases}
	\frac{ A \lambda_2 \sqrt{A \lambda_2} e^{-\frac{u_2}{A \lambda_2 \rho}}}{(B u_2 \lambda_1 + A \lambda_2)(\lambda_{1t}-\lambda_{2t}) \sqrt{u_2 \eta \lambda_{tb}}}

\sum\limits_{i=1}^{2} (-1)^{i+1} \sqrt{ \lambda_{it}  } e^{\frac{A\lambda_2}{2 \eta u_2 \lambda_{it}\lambda_{tb} }  }  W_{-\frac{1}{2},0}\left( \frac{A\lambda_2}{ \eta u_2\lambda_{it}\lambda_{tb}}\right), & \lambda_{1t} \ne \lambda_{2t}
		\\

\frac{(A \lambda_2)^2 e^{-\frac{u_2}{A \lambda_2 \rho}}}{\eta u_2 \lambda \lambda_{tb}(B u_2 \lambda_1 + A \lambda_2)} 	
 e^{\frac{A\lambda_2}{2 u_2 \eta \lambda \lambda_{tb}}} W_{-1,-\frac{1}{2}}\left(\frac{A\lambda_2}{u_2 \eta \lambda \lambda_{tb}}\right) , & \lambda_{1t} = \lambda_{2t} = \lambda
	\end{cases}  \label{I_11_v1}
\end{equation}

\hrulefill
\vspace{-0.5cm}
\end{figure*}

\end{thm}
\begin{IEEEproof}
  Applying (\ref{gamma_x2}) in (\ref{OP_2}) and exploiting the law of total probability, we get  
 \begin{equation}
	\begin{split}
		P_{2}^{o} &\!=\! \Pr\left(\!\underset{i \in \{1,2\}}{\text{rand}}{U_i}\!=\!U_1\!\right)\! \tilde{P}_{2}^{o}|_{\epsilon \!= \!0} \!+\! \Pr\left(\!\underset{i \in \{1,2\}}{\text{rand}}{U_i}\!=\!U_2\!\right) \! \tilde{P}_{2}^{o}|_{\epsilon \!= \!1} ,   \label{OP_x2_v1} 
	\end{split}	
\end{equation}
where
\begin{equation}
	\tilde{P}_{2}^{o}=\Pr \left( \frac{A \rho |h_2|^2}{B \rho |h_1|^2 + \eta \rho Z   + 1} < u_2  \right) \label{P2_v0}
\end{equation}
and $Z\!=\!W |h_{tb}|^2$ with $W\!=\!|h_{2t}|^2\!+\!|h_{1t}|^2$. Given that $|h_{1}|^2$ and $|h_{2}|^2$ are exponentially distributed random variables (RVs), their probability density function (PDF) and cumulative density function (CDF) can be given as $f_{h_n}(x)=\frac{1}{\lambda_n}e^{-\frac{x}{\lambda_n}}$ and $F_{h_n}(x)\!=\!1\!-\!e^{-\frac{x}{\lambda_n}}$, respectively, with $n\!\in\! \{1,2\}$. Furthermore, $W$ occurs as the sum of two independent exponentially distributed RVs; thus, its PDF can be given as \cite{karagiannidis2006closed}
\begin{equation}
	 f_{W}(w)=\begin{cases}
	 	\frac{1}{\lambda_{1t}-\lambda_{2t}} ( e^{-\frac{w}{\lambda_{1t}}} - e^{-\frac{w}{\lambda_{2t}}} ), & \lambda_{1t} \ne \lambda_{2t}
	 	\\
	 	\\
	 	\frac{1}{\lambda^2} w e^{-\frac{w}{\lambda}}, & \lambda_{1t} = \lambda_{2t}=\lambda.
	 \end{cases} \label{f_W}
\end{equation}

Next, the PDF of RV $Z$ is going to be evaluated by firstly extracting RV's $Z$ CDF and then taking its derivative. Via the definition of the CDF,
 it holds\begin{equation}
	\begin{split}
	F_{Z}(z)&\!=\!\Pr(W |h_{tb}|^2\!<\!z) \!=\! \int_{0}^{\infty} F_{|h_{tb}|^2}\left(\frac{z}{w}\right) f_{W}(w) \mathrm{d}w.
	\end{split} \label{F_Z}
\end{equation}
Focusing on the case when $\lambda_{1t} \ne \lambda_{2t}$, invoking the first branch of (\ref{f_W}) into (\ref{F_Z}), it holds
\begin{equation}
	\begin{split}
		F_{Z}(z)&= 1 - 	\frac{1}{\lambda_{1t}-\lambda_{2t}} \left( \int_{0}^{\infty} e^{-\frac{w}{\lambda_{1t}}-\frac{z}{\lambda_{tb}w}} \mathrm{d}w \right. \\& \left.
		 \ \ \ \ \ \ \ \ \ \ \ \ \ \ \ \ - \int_{0}^{\infty} e^{-\frac{w}{\lambda_{2t}}-\frac{z}{\lambda_{tb}w}} \mathrm{d}w \right).
	\end{split} \label{F_Z_v2}
\end{equation}
Utilizing \cite[(3.324.1)]{grad}, (\ref{F_Z_v2}) turns into
\begin{equation}
	\begin{split}
		F_{Z}(z)&= 1 - 	\frac{1}{\lambda_{1t}-\lambda_{2t}} \left( 2 \sqrt{\frac{z\lambda_{1t}}{\lambda_{tb}}} K_1\left(2 \sqrt{\frac{z}{\lambda_{1t}\lambda_{tb}}}\right) \right. \\& \left.
\ \ \ \ \ \ \ \ \  \ \ \ \ \ - 2 \sqrt{\frac{z\lambda_{2t}}{\lambda_{tb}}} K_1\left(2 \sqrt{\frac{z}{\lambda_{2t}\lambda_{tb}}}\right) \right),
\end{split} \label{F_Z_v3}
\end{equation} 
where $K_v(\cdot)$ denotes the modified Bessel function of the second kind with order $v$. On the other hand, when $\lambda_{1t} \!=\! \lambda_{2t}\!=\!\lambda$, by invoking the second branch of (\ref{f_W}), (\ref{F_Z}) becomes 
\begin{equation}
	\begin{split}
	F_{Z}(z)&= 1 - \frac{1}{\lambda^2} \int_{0}^{\infty}  w e^{-\frac{w}{\lambda}-\frac{z}{\lambda_{tb}w}} 	\mathrm{d}w.
\end{split} \label{F_Z_v4}
\end{equation}
Applying \cite[(3.471.9)]{grad}, (\ref{F_Z_v4}) is transformed into
\begin{equation}
	\begin{split}
		F_{Z}(z)&= 1 -  \frac{2 z}{\lambda \lambda_{tb}}K_2\left(2 \sqrt{\frac{z}{\lambda\lambda_{tb}}}\right) .
	\end{split} \label{F_Z_v5}
\end{equation}
By taking the derivatives of (\ref{F_Z_v3}) and (\ref{F_Z_v5}) and exploiting Bessel functions' properties \cite[(03.04.7.0002.01), (03.04.20.0006.01)]{wolf}, we get RV's $Z$ PDF as follows
\begin{equation}
	f_{Z}(z)\!=\!\begin{cases}
		\!\frac{2}{(\lambda_{1t}\!-\!\lambda_{2t})\lambda_{tb}} \!\sum\limits_{i=1}^{2} \!(\!-1)^{i+1} \! K_0\!\left(\!2 \sqrt{\!\frac{z}{\lambda_{it}\lambda_{tb}}\!}\!\right)\!, &\! \lambda_{1t} \ne \lambda_{2t}
		\\
		\! \frac{2}{\lambda \lambda_{tb}} \sqrt{\frac{z}{\lambda \lambda_{tb}}} K_1\left(2 \sqrt{\frac{z}{\lambda \lambda_{tb}}}\right)\!,\! & \lambda_{1t} \!=\! \lambda_{2t} \!=\! \lambda.
	\end{cases} \label{F_Z_v6}
\end{equation}

Having obtained RV's $Z$ PDF, we are able to further proceed into the extraction of $U_2$'s OP. Considering that $|h_{1}|^2$, $|h_{2}|^2$ and $Z$ are independent with each other, (\ref{P2_v0}) can be written as
\begin{equation}
	\begin{split}
	\tilde{P}_{2}^{o} &= \int_{0}^{\infty}   \int_{0}^{\infty}  \int_{0}^{ \frac{  B \rho u_2 y + \eta \rho z u_2  + u_2}{A \rho} } f_{|h_{2}|^2}(x) f_{|h_{1}|^2}(y) \\
	& \ \ \ \ \ \ \ \ \ \ \ \ \ \ \ \ \ \ \ \ \ \ \ \ \ \ \ \ \ \ \ \ \ \ \ \ \  
	\times f_{Z}(z) \, \mathrm{d}x \, \mathrm{d}y \, \mathrm{d}z \\
	&=1- {I_1},  	
	\end{split} \label{P2_v2}
\end{equation}
where
\begin{align}
I_1 = \int_{0}^{\infty} \int_{0}^{\infty} e^{ -\frac{  B \rho u_2 y + \eta \rho z u_2  + u_2}{A \lambda_2 \rho} } f_{|h_{1}|^2}(y) f_{Z}(z) \mathrm{d}y \mathrm{d}z.
\label{Eq:I1}
\end{align}
	Substituting $f_{|h_{1}|^2}(y)$, \eqref{Eq:I1} becomes
\begin{equation}
	\begin{split}
	I_1 &= \frac{1}{\lambda_1} e^{-\frac{u_2}{A \lambda_2 \rho}} \int_{0}^{\infty} f_Z(z) e^{-\frac{u_2 \eta z}{A \lambda_2}} \mathrm{d}z  \int_{0}^{\infty}  e^{-\frac{y}{\lambda_1}}  e^{-\frac{B u_2  y}{A \lambda_2}} \mathrm{d}y  \\
	&=  \frac{A \lambda_2 e^{-\frac{u_2}{A \lambda_2 \rho}}}{B u_2 \lambda_1 + A \lambda_2} \int_{0}^{\infty} f_Z(z) e^{-\frac{u_2 \eta z}{A \lambda_2}} \mathrm{d}z.   
	\end{split} \label{P2_v3}
\end{equation}
Considering the case $\lambda_{1t} \!\ne\! \lambda_{2t}$, by invoking the first branch of (\ref{f_W}) into $I_{1}$, it yields
\begin{equation}
	\begin{split}		
	I_{1}&= \frac{2 A \lambda_2 e^{-\frac{u_2}{A \lambda_2 \rho}}}{\lambda_{tb}(\lambda_{1t}-\lambda_{2t})(B u_2 \lambda_1 + A \lambda_2)}  \left( \int_{0}^{\infty}  K_0\left(2 \sqrt{\frac{z}{\lambda_{1t}\lambda_{tb}}}\right) \right. \\& \left.	\times e^{-\frac{u_2 \eta z}{A \lambda_2}} \mathrm{d}z 	- \int_{0}^{\infty} K_0\left(2 \sqrt{\frac{z}{\lambda_{2t}\lambda_{tb}}}\right) e^{-\frac{u_2 \eta z}{A \lambda_2}} \mathrm{d}z \right). 
	\end{split} \label{P2_v4}
\end{equation} 
On the other hand, when $\lambda_{1t} \!=\! \lambda_{2t}\!=\!\lambda$, $I_{1}$ can be written as
\begin{equation}
	\begin{split}
		I_{1}\!=\! \frac{2 A \lambda_2 e^{-\frac{u_2}{A \lambda_2 \rho}}}{\lambda \lambda_{tb}(B u_2 \lambda_1 \!+\! A \lambda_2)}\! \int_{0}^{\infty} \!\sqrt{\!\frac{z}{\lambda \lambda_{tb}}} &K_1\!\left(\!2 \sqrt{\frac{z}{\lambda \lambda_{tb}}}\!\right)\!  e^{-\frac{u_2 \eta z}{A \lambda_2}} \mathrm{d}z.  
	\end{split} \label{P2_v6}
\end{equation} 
By applying \cite[(6.614.4)]{grad} in (\ref{P2_v4}) as well as by changing variable of $x\!=\!\frac{z}{\lambda \lambda{tb}}$ and then applying \cite[(6.643.6)]{grad} in (\ref{P2_v6}), $I_{1}$ can be calculated from (\ref{I_11_v1}) as given at the top of the page.
By invoking (\ref{I_11_v1}) in (\ref{P2_v2}) and then the resultant (\ref{P2_v2}) in (\ref{OP_x2_v1}), we get (\ref{x2_OP}) and this concludes the proof.     
\end{IEEEproof}



\vspace{-0.2cm}
\subsection{Outage performance of $U_{1}$}

It is noted that in order to avoid an outage for $U_1$, BS should firstly decode $x_2$ and then obtain $x_1$ via SIC. Hence, The OP of $U_1$ can be expressed as 
\begin{equation}
	P_{1}^{o} = 1 - \Pr \left( \gamma_{x_2} > u_2, \gamma_{x_1} > u_1   \right), \label{P1_v1}	
\end{equation}
where $u_1\!=\!2^{R_{1}}-1$. The following theorem provides $U_{1}$'s OP.
\begin{thm} \label{thm_U1}
The OP of $U_1$, under ipSIC, can be evaluated as 
\begin{equation}
	P_{1}^{o}\!=\! \begin{cases}
	     \!1 \!-\! \frac{1}{2} \left( I_{2}|_{\epsilon = 0} \!-\! I_{3}|_{\epsilon = 0}    \!+\! I_{2}|_{\epsilon = 1} \!- \!I_{3}|_{\epsilon = 1}  \right), & k_2 u_2 u_1 \!<\!1 
      \\
      \!1, & \text{otherwise},
      \end{cases}  \label{x1_OP_v0}
\end{equation}
where $I_2$, $I_3$ are given at the top of the next page via (\ref{I2_}) and (\ref{I_3_f}), respectively, with
\begin{figure*} \vspace{-0.5cm}
\begin{equation}
	I_{2}=\begin{cases}
		\frac{ A \rho k_2  \lambda_2  u_1  e^{\frac{ 1}{A \rho k_2 \lambda_2}  -  \frac{1}{A C \rho}(\frac{1}{\lambda_1} + \frac{B  }{A  k_2 \lambda_2 u_1}) (u_2  + \frac{1}{ k_2} )}}{  \sqrt{\lambda_{tb}Q_1} (\lambda_{1t}-\lambda_{2t})  (A \rho k_2 \lambda_2 u_1+ B \rho \lambda_1)}   \sum\limits_{i=1}^{2} (-1)^{i+1} \sqrt{ \lambda_{it}} e^{\frac{1}{2 Q_1 \lambda_{it} \lambda_{tb}}} W_{-\frac{1}{2},0} \left(\frac{1}{ Q_1 \lambda_{it} \lambda_{tb}}\right), & \lambda_{1t} \ne \lambda_{2t}
		\\
		\frac{ A \rho k_2  \lambda_2 u_1  e^{\frac{ 1}{A \rho k_2 \lambda_2 }  -  \frac{1}{A C \rho}(\frac{1}{\lambda_1} + \frac{B  }{A  k_2 \lambda_2 u_1}) (u_2  + \frac{1}{ k_2} )}}{ Q_1 \lambda \lambda_{tb} (A \rho k_2 \lambda_2 u_1+ B \rho \lambda_1) } e^{\frac{1}{2 Q_1 \lambda \lambda_{tb}}} W_{-1,-\frac{1}{2}} \left(\frac{1}{ Q_1 \lambda \lambda_{tb}}\right) , & \lambda_{1t} = \lambda_{2t} = \lambda
	\end{cases}  \label{I2_} \vspace{-0.15cm}
\end{equation}
\vspace{-0.3cm}
\hrulefill
\end{figure*}
\begin{figure*}
\begin{equation}
	I_{3}=\begin{cases}
		\frac{ A \lambda_2   e^{-\frac{  u_2  }{ A \lambda_2 \rho}  -  (\frac{B u_2}{A \lambda_2}+\frac{1}{\lambda_1})(\frac{u_2}{A C \rho} + \frac{1}{A C k_2 \rho})} }{  \sqrt{\lambda_{tb} Q_2}  (\lambda_{1t}-\lambda_{2t}) (B u_2 \lambda_1  +  A \lambda_2)  }   \sum\limits_{i=1}^{2} (-1)^{i+1} \sqrt{ \lambda_{it}} e^{\frac{1}{2 Q_2 \lambda_{it} \lambda_{tb}}} W_{-\frac{1}{2},0} \left(\frac{1}{ Q_2 \lambda_{it} \lambda_{tb}}\right), & \lambda_{1t} \ne \lambda_{2t}
		\\
		\frac{ A \lambda_2  e^{-\frac{  u_2  }{ A \lambda_2 \rho}  -  (\frac{B u_2}{A \lambda_2}+\frac{1}{\lambda_1})(\frac{u_2}{A C \rho} + \frac{1}{A C k_2 \rho})} }{  Q_2 \lambda \lambda_{tb} (B u_2 \lambda_1  +  A \lambda_2)} e^{\frac{1}{2 Q_2 \lambda \lambda_{tb}}} W_{-1,-\frac{1}{2}} \left(\frac{1}{ Q_2 \lambda \lambda_{tb}}\right) , & \lambda_{1t} = \lambda_{2t} = \lambda
	\end{cases}  \label{I_3_f} \vspace{-0.15cm}
\end{equation}
\hrulefill
\vspace{-0.6cm}
\end{figure*}
    \begin{subequations}   
\begin{align}
      C=&\frac{B}{A k_2 u_1}-\frac{B u_2}{A}, 
 \label{C}
\\
    Q_1 =  (\frac{1}{\lambda_1} + \frac{B }{A k_2 \lambda_2 u_1})& (\frac{ \eta }{A C k_2} + \frac{\eta  u_2}{A C}) - \frac{ \eta }{A  k_2 \lambda_2 }, \label{Q1}
\\
Q_2 = \frac{ \eta   u_2    }{ A \lambda_2 }  + (\frac{B u_2}{A \lambda_2}&+\frac{1}{\lambda_1})(\frac{\eta}{A C k_2}+\frac{\eta u_2}{A C}).   \label{Q2}  
\end{align}
\end{subequations}

\end{thm}

\begin{IEEEproof}
 Applying (\ref{gamma_x2}) and (\ref{gamma_x1}) in (\ref{P1_v1}) and exploiting the law of total probability, $U_1$'s OP can be expressed as 
 \begin{equation}
	\begin{split}
		P_{1}^{o}&\!=\! \Pr\left(\!\underset{i \in \{1,2\}}{\text{rand}}{U_i}\!=\!U_1\!\right)\! \tilde{P}_{1}^{o}|_{\epsilon \!=\! 0} \!+\! \Pr\left(\!\underset{i \in \{1,2\}}{\text{rand}}{U_i}\!=\!U_2\!\right) \! \tilde{P}_{1}^{o}|_{\epsilon \!=\! 1},   \label{OP_x1_v1} 
	\end{split}	
\end{equation}
where 
\begin{equation}
    \begin{split}
     \tilde{P}_{1}^{o} &= 1 - \Pr \left(   |h_2|^2 > \frac{B \rho |h_1|^2 u_2 + \eta \rho Z u_2  +  u_2  }{A \rho}, \right. \\& \left. \ \ \ \  \ \ \ \  \ \ \ \ \ \  \ \ \ |h_2|^2 < \frac{B \rho |h_1|^2 - \eta \rho Z u_1 - u_1}{A \rho k_2 u_1}   \right).   
    \end{split}  \label{P1_v2}  
\end{equation}
To ensure that the probability that appears in (\ref{P1_v2}) takes non-zero values, the following must hold

\begin{equation}
    \begin{split}
     \frac{B \rho |h_1|^2 u_2 + \eta \rho Z u_2  +  u_2  }{A \rho} < \frac{B \rho |h_1|^2 - \eta \rho Z u_1 - u_1}{A \rho k_2 u_1}      
    \end{split}  \label{P1_v3}  
\end{equation}
or equivalently
\begin{equation}
    \begin{split}
      C \, |h_1|^2 > \frac{ \eta Z}{A k_2} + \frac{\eta Z u_2}{A} + \frac{u_2}{A \rho}  + \frac{1}{A \rho k_2}, 
    \end{split}  \label{P1_v4}  
\end{equation}
where $C$ is provided in (\ref{C}). It can be observed that when $C<0$, then (\ref{P1_v4}) does not hold. On the other hand, when $C>0$, i.e., $k_2 u_2 u_1 < 1$, then (\ref{P1_v4}) holds only when

\begin{equation}
    \begin{split}
     |h_1|^2 > \frac{1}{C} \left( \frac{ \eta Z}{A k_2} + \frac{\eta Z u_2}{A} + \frac{u_2}{A \rho}  + \frac{1}{A \rho k_2}\right) \overset{\triangle}{=}D_z.     
    \end{split}  \label{P1_v5}  
\end{equation}
Focusing on the $C>0$ case and taking into account (\ref{P1_v5}), (\ref{P1_v2}) can be rewritten as 
\begin{equation}
\begin{split}
    \tilde{P}_{1}^{o} &= 1 -
        \int_{0}^{\infty} \int_{D_z}^{\infty} \left( F_{|h_2|^2} \left( \frac{B \rho y - \eta \rho z u_1 - u_1}{A \rho k_2 u_1}  \right)  \right. \\& \left.
        -   F_{|h_2|^2} \left( \frac{B \rho y u_2 + \eta \rho z u_2  +  u_2  }{A \rho}  \right) \right)  f_{|h_1|^2}(y) f_Z(z) \mathrm{d}y \mathrm{d}z. 
\end{split}   \label{P1_v6}  
\end{equation}
By Substituting $F_{|h_2|^2}$ and performing some algebraic manipulations, (\ref{P1_v6}) can be transformed into
\begin{equation}
   \tilde{P}_{1}^{o} = 1 + I_2 - I_3, \label{P1_tilde}
\end{equation}
where
\begin{equation}
\begin{split}
    I_2 &=
        \int_{0}^{\infty} f_Z(z) \, \mathrm{d}z \int_{D_z}^{\infty}  e^{-\frac{B \rho y - \eta \rho z u_1 - u_1}{A \rho k_2 \lambda_2 u_1}}  
            f_{|h_1|^2}(y) \, \mathrm{d}y, 
\end{split}  \label{I_2}  
\end{equation}
\begin{equation}
\begin{split}
    I_3 &=
        \int_{0}^{\infty} f_Z(z) \, \mathrm{d}z \int_{D_z}^{\infty}  e^{-\frac{B \rho y u_2 + \eta \rho z u_2  +  u_2  }{ A \lambda_2 \rho}}  
            f_{|h_1|^2}(y) \, \mathrm{d}y.
\end{split}  \label{I_3}  
\end{equation}
Reagrding $I_2$, it can be equivalently rewritten as
\begin{equation}
\begin{split}
    I_2 &= e^{\frac{ u_1}{A \rho k_2 \lambda_2 u_1}}
        \int_{0}^{\infty} e^{ \frac{ \eta \rho z u_1}{A \rho k_2 \lambda_2 u_1}} f_Z(z) \, \mathrm{d}z \\
        & \hspace{-0.2cm} \times \int_{D_z}^{\infty}  e^{-\frac{B \rho y }{A \rho k_2 \lambda_2 u_1}}  
            f_{|h_1|^2}(y) \, \mathrm{d}y. 
\end{split}  \label{I_2}  
\end{equation}
By substituting $f_{|h_1|^2}(y)$ and performing an exponential integration, yields
\begin{equation}
\begin{split}
    I_2 &=  \frac{ A \rho k_2  \lambda_2 u_1  e^{\frac{ u_1}{A \rho k_2 \lambda_2 u_1}  -  \frac{1}{A C \rho}(\frac{1}{\lambda_1} + \frac{B \rho }{A \rho k_2 \lambda_2 u_1}) (u_2  + \frac{1}{ k_2} )}}{A \rho k_2 \lambda_2 u_1+ B \rho \lambda_1} \\
    & \times
        \int_{0}^{\infty} e^{ - Q_1 z}   f_Z(z) \, \mathrm{d}z, 
\end{split}  \label{I_2_v2}  
\end{equation}
where $Q_1$ is given via (\ref{Q1}). 

In a similar manner with $I_2$, $I_3$ can be rewritten as
\begin{equation}
\begin{split}
    I_3 &= e^{-\frac{  u_2  }{ A \lambda_2 \rho}}
        \int_{0}^{\infty} f_Z(z) e^{-\frac{ \eta \rho z u_2    }{ A \lambda_2 \rho}} \, \mathrm{d}z \int_{D_z}^{\infty}  e^{-\frac{B \rho y u_2  }{ A \lambda_2 \rho}}  
            f_{|h_1|^2}(y) \, \mathrm{d}y,
\end{split}  \label{I_3}  
\end{equation}
which, after substituting $f_{|h_1|^2}(y)$, becomes
\begin{equation}
\begin{split}
    I_3 &\!=\! \frac{ A \lambda_2  e^{-\frac{  u_2  }{ A \lambda_2 \rho} \! -\!  (\frac{B u_2}{A \lambda_2}\!+\!\frac{1}{\lambda_1})(\frac{u_2}{A C \rho} \!+\! \frac{1}{A C k_2 \rho})} }{B u_2 \lambda_1  +  A \lambda_2}\!     \int_{0}^{\infty} f_Z(z)  e^{- Q_2 z  } \, \mathrm{d}z, 
\end{split}  \label{I_3_v1}  
\end{equation}
where $Q_2$ is given via (\ref{Q2}).

At this point it is noted that the integrals of (\ref{I_2_v2}) and (\ref{I_3_v1}) can be evaluated if $Q_1>0$, $Q_2>0$, respectively. Indeed, it can be straightforwardly shown that $Q_1,Q_2>0$. Thus, for the case when $\lambda_{1t} \!\ne\! \lambda_{2t}$ by substituting the first branch of (\ref{F_Z_v6}) into (\ref{I_2_v2}), (\ref{I_3_v1}) and utilizing \cite[(6.614.4)]{grad} we get the first branch of (\ref{I2_}) and (\ref{I_3_f}), respectively. On the other hand, for the case when $\lambda_{1t} \!=\! \lambda_{2t}$, by substituting the second branch of (\ref{F_Z_v6}) into (\ref{I_2_v2}), (\ref{I_3_v1}) and leveraging \cite[(6.643.6)]{grad} we get the second branch of (\ref{I2_}) and (\ref{I_3_f}), respectively. Applying (\ref{I2_}) and (\ref{I_3_f}) in (\ref{P1_tilde}) and then the resultant (\ref{P1_tilde}) in (\ref{OP_x1_v1}) and taking into account the condition $C\!>\!0$, then (\ref{x1_OP_v0}) is derived. This concludes the proof.


\end{IEEEproof}

It is noted that the above extracted expression holds for the ipSIC case. Inconveniently, similarly with other cases in the existing literature, $U_1$'s OP under pSIC cannot be obtained by just setting $k_1\!=\!k_2\!=\!0$ in (\ref{x1_OP_v0}); thus, in what follows we provide a lemma that returs the OP of $U_1$ under the ideal case of pSIC.  

\begin{lem}
    The OP of $U_1$ under pSIC can be calculated as
    \begin{equation}
	P_{1}^{p}=   1 - \frac{1}{2} \left( I_{1}^p|_{\epsilon = 0} + I_{1}^p|_{\epsilon = 1}     \right),   \label{x1_OP_pSIC}
\end{equation}
where $I_{1}^p$ is given at the top of the next page via (\ref{I1_p}), where
\begin{equation}
    Q_{1}^{p} = \frac{\eta u_2}{A \lambda_2} + \left(\frac{1}{\lambda_1}+\frac{B u_2}{A \lambda_2}\right) \frac{\eta u_1}{B}.   \label{Q_11}
\end{equation}

\end{lem}

\begin{IEEEproof}
It is noted that (\ref{OP_x1_v1}) still holds as soon as $\tilde{P}_{1}^{o}$ is properly updated. Specifically, by setting $k_1=k_2=0$ in (\ref{gamma_x2}), (\ref{gamma_x1}) in order to correspond to the pSIC case, $\tilde{P}_{1}^{o}$ is updated as
\begin{equation}
    \begin{split}
     \tilde{P}_{1}^{p} &= 1 - \Pr \left(   |h_2|^2 > \frac{B \rho |h_1|^2 u_2 + \eta \rho Z u_2  +  u_2  }{A \rho}, \right. \\& \left. \ \ \ \  \ \ \ \  \ \ \ \ \ \  \ \ \  |h_1|^2 > \frac{\eta \rho Z u_1 + u_1}{B \rho}   \right) \\
     & =  1 - I_1^p, 
    \end{split}  \label{P1_psic_v1}  
\end{equation}
where 
\begin{equation}
\begin{split}
    I_1^p &= \int_{0}^{\infty} \int_{\frac{\eta \rho z u_1 + u_1}{B \rho}}^{\infty} \int_{\frac{B \rho y u_2 + \eta \rho z u_2  +  u_2  }{A \rho}}^{\infty} f_{|h_{2}|^2}(x) f_{|h_{1}|^2}(y) f_{Z}(z) \\
     & \ \ \ \ \ \ \ \ \ \ \ \ \ \ \ \ \  \ \ \ 
 \ \ \ \  \ \ \ \ \ \ \ \ \ \  \ \ \ \ \ \ \ \  \ \ \ \ \ \ \ \ \  \times \mathrm{d}x \mathrm{d}y \mathrm{d}z \\
 & =  e^{-\frac{u_2}{A \lambda_2 \rho}} \! \int_{0}^{\infty}\! e^{-\frac{ \eta u_2 z}{A \lambda_2}}\! f_{Z}(z)\mathrm{d}z\!  \int_{\frac{\eta \rho z u_1 \!+\! u_1}{B \rho}}^{\infty}\! e^{-\frac{u_2 B y}{A \lambda_2}} f_{|h_{1}|^2}(y) \mathrm{d}y  \\
 & \overset{(s_1)}{=} \frac{A \lambda_2 e^{-\frac{u_2}{A \lambda_2 \rho}- (\frac{1}{\lambda_1}\!+\!\frac{B u_2}{ A \lambda_2}) \frac{u_1}{B \rho}}}{A \lambda_2 \!+\! B u_2 \lambda_1} \int_{0}^{\infty}\! e^{-Q_{1}^{p} z} f_{Z}(z) \mathrm{d}z,
    \end{split} \label{P_1_p}
\end{equation}
where step ($s_1$) occurs by invoking $f_{|h_{1}|^2}(y)$ and $Q_{1}^{p}$ is given via (\ref{Q_11}). At this point it is noted that the integral of (\ref{P_1_p}) can be evaluated if $Q_{1}^{p}\!>\!0$. Indeed, it can be straightforwardly shown that $Q_{1}^{p}\!>\!0$. Thus, for the case when $\lambda_{1t} \! \ne \! \lambda_{2t}$ by substituting the first branch of (\ref{F_Z_v6}) into (\ref{P_1_p}) and utilizing \cite[(6.614.4)]{grad}, the first branch of (\ref{I1_p}) occurs. On the other hand, for the case when $\lambda_{1t} \!=\! \lambda_{2t}$, by substituting the second branch of (\ref{I1_p}) into (\ref{P_1_p}) and leveraging \cite[(6.643.6)]{grad}, we get the second branch of (\ref{I1_p}). By applying (\ref{I1_p}) in (\ref{P1_psic_v1}) and then the resultant (\ref{P1_psic_v1}) in (\ref{OP_x1_v1}), we get (\ref{x1_OP_pSIC}) and this concludes the proof.

\begin{figure*} \vspace{-0.3cm}
\begin{equation}
	I_{1}^{p}=\begin{cases}
		\frac{A \lambda_2 e^{-\frac{u_2}{A \lambda_2 \rho}- (\frac{1}{\lambda_1}+\frac{B u_2}{ A \lambda_2}) \frac{u_1}{B \rho}}}{\sqrt{Q_{1}^{p} \lambda_{tb}}(\lambda_{1t}-\lambda_{2t})(A \lambda_2 + B u_2 \lambda_1)}   \sum\limits_{i=1}^{2} (-1)^{i+1} \sqrt{ \lambda_{it}} e^{\frac{1}{2 Q_{1}^{p} \lambda_{it} \lambda_{tb}}} W_{-\frac{1}{2},0} \left(\frac{1}{ Q_{1}^{p} \lambda_{it} \lambda_{tb}}\right), & \lambda_{1t} \!\ne\! \lambda_{2t}
		\\
		\frac{A \lambda_2 e^{-\frac{u_2}{A \lambda_2 \rho}- (\frac{1}{\lambda_1}+\frac{B u_2}{ A \lambda_2}) \frac{u_1}{B \rho}}}{Q_{1}^{p} \lambda \lambda_{tb}(A \lambda_2 + B u_2 \lambda_1)} e^{\frac{1}{2 Q_{1}^{p} \lambda \lambda_{tb}}} W_{-1,-\frac{1}{2}} \left(\frac{1}{ Q_{1}^{p} \lambda \lambda_{tb}}\right) , & \lambda_{1t} \!=\! \lambda_{2t} \!=\! \lambda.
	\end{cases}  \label{I1_p}  \vspace{-0.15cm}
\end{equation}
\hrulefill
\end{figure*}   
\end{IEEEproof}

\vspace{-0.4cm}
\subsection{Outage performance of BD}

Before proceeding into the extraction of BD's OP, a useful lemma is provided and proved. Specifically, in this lemma, an integral that is crucial for the conducted performance analysis is solved in an exact closed-form manner via the help of the extended generalized bivariate Fox-H function (EGBFHF) defined in \cite[(2.57)]{mathai2009h}. It is noted that although the EGBFHF has not been included as a built-in function in the widely used mathematical softwares, in the existing literature there have been provided custom Matlab, Python as well as Mathematica codes for its implementation \cite{peppas2012new,lei2017secrecy,alhennawi2015closed}. However, it is also important to mention that the use of these existing codes demands the careful selection of the integration contours which also may change for different parameter selection as well as it may lead to enhanced computational complexity. Hence, considering the above, except from the exact solution, we also provide a simple approximation for the considered integral via the well-known Gauss-Chebyshev quadrature which offers the advantage of controllable computational complexity.    

\begin{lem} \label{lem:G}
Let $\alpha$, $\beta$ being positive constants, then the integral
    \begin{equation}
       \Phi(\alpha, \beta) \overset{\triangle}{=} \int_{\alpha}^{\infty} e^{- \beta z} f_{Z}(z) \mathrm{d}z  \label{G_def}
    \end{equation}
can be calculated in an exact closed-form manner via (\ref{G_ex}) given at the top of the next page, where $H^{m_1,n_1:m_2,n_2:m_3,n_3}_{p_1,q_1:p_2,q_2:m_3,q_3}(\cdot)$ denotes the EGBFHF. Furthermore, a useful approximation of (\ref{G_def}) is given via (\ref{G_apx}), where $\Delta_l$ is a complexity-accuracy trade-off parameter and $\psi_j\!=\! cos(\frac{\pi (2j-1)}{2 \Delta_l})$ with $l=1,2,3$. 

\begin{figure*} \vspace{-0.45cm}
\begin{equation}
\Phi(\alpha, \beta) \!=\! \begin{cases}
\frac{2}{\lambda_{tb}(\lambda_{1t}-\lambda_{2t})} \sum_{i=1}^{2} (-1)^{i+1} \lambda_{it} \, \lambda_{tb} H^{0,2:1,0:0,1}_{2,0:0,1:1,1}\! \left(^{(0;1,1),(0;1,1)}_{\ \ \ \ \ \ \ -} \Big| \, ^{\ ^-}_{(0,1)} \Big| \, ^{(1,1)}_{(0,1)} \Big| \beta \lambda_{it} \lambda_{tb} , \frac{\lambda_{it} \lambda_{tb}}{\alpha}  \right)  , & \lambda_{1t} \ne \lambda_{2t}
\\
2 \, H^{0,2:1,0:0,1}_{2,0:0,1:1,1}\! \left(^{(0;1,1),(-1;1,1)}_{\ \ \ \ \ \ \ -} \Big| \, ^{\ ^-}_{(0,1)} \Big| \, ^{(1,1)}_{(0,1)} \Big| \beta \lambda \lambda_{tb} , \frac{\lambda \lambda_{tb}}{\alpha}  \right)
, & \lambda_{1t} \!=\! \lambda_{2t} \!=\! \lambda.
\end{cases}   \label{G_ex} \vspace{-0.15cm}
\end{equation}
\hrulefill
\vspace{-0.5cm}
\end{figure*}

\begin{figure*} \vspace{-0.1cm}
\begin{equation}
\Phi(\alpha, \beta) \!\approx\! \begin{cases}
\frac{1}{\sqrt{\beta  \lambda_{tb}} (\lambda_{1t} - \lambda_{2t})} \sum\limits_{i=1}^{2} (-1)^{i+1} \sqrt{\lambda_{it}} e^{\frac{1}{2 \beta \lambda_{it} \lambda_{tb}}}  W_{-\frac{1}{2},0} \left( \frac{1}{ \beta \lambda_{it} \lambda_{tb}} \right)  -   \frac{\pi \alpha}{\lambda_{tb}(\lambda_{1t}-\lambda_{2t})} \sum\limits_{i=1}^{2} \frac{(-1)^{i+1}}{\Delta_i} \\
\times \sum\limits_{j=1}^{\Delta_i} \sqrt{1-\psi_j^2} K_0\left( 2 \sqrt{\frac{\alpha (\psi_j+1)}{2 \lambda_{it} \lambda_{tb}}} \right) e^{-\frac{\alpha \beta (\psi_j+1)}{2}}    , & \lambda_{1t} \ne \lambda_{2t}
\\
\frac{1}{\beta \lambda \lambda_{tb}} e^{\frac{1}{2 \beta \lambda \lambda_{tb}}} W_{-1,-\frac{1}{2}} \left( \frac{1}{ \beta \lambda_{it} \lambda_{tb}} \right)     -     \frac{\pi \alpha}{\lambda \lambda_{tb} \Delta_3} \sum\limits_{j=1}^{\Delta_3} \sqrt{\frac{\alpha (1+\psi_j) (1-\psi_j^2)}{2 \lambda \lambda_{tb}}}  K_1 \left(2 \sqrt{\frac{\alpha (\psi_j+1)}{2 \lambda \lambda_{tb}}} \right)  e^{-\frac{\alpha \beta (\psi_j+1)}{2}} 
, & \lambda_{1t} \!=\! \lambda_{2t} \!=\! \lambda.
\end{cases}   \label{G_apx} \vspace{-0.15cm}
\end{equation}
\vspace{-0.3cm}
\vspace{-0.4cm}
\end{figure*}

    \label{lem:G}
\end{lem}

\begin{IEEEproof} The proof is provided in Appendix A.

\end{IEEEproof}

In what follow, we focus on the calculation of BD's OP. In a similar manner with the decoding mechanism followed for $U_1$, in order to decode $x_t$, the BS should successfully decode $x_2$ and $x_1$ and then attempt to obtain $x_t$ via SIC. Hence, the OP of the BD can be expressed as 
\begin{equation}
	P_{t}^{o} = 1 - \Pr \left( \gamma_{x_2} > u_2, \gamma_{x_1} > u_1, \gamma_{x_t} > u_t   \right), \label{Pt_v1}	
\end{equation}
where $u_t\!=\!2^{R_{t}}-1$.
Next, a theorem about the OP of the BD is presented.
\begin{thm} \label{thm:OP_BD}
The OP of the BD under ipSIC can be given by
\begin{align}
	P_{t}^{o}&=   1 + \frac{1}{2} \left( P_{t,11}|_{\epsilon = 0} - P_{t,12}|_{\epsilon = 0} - P_{t,21}|_{\epsilon = 0} + P_{t,22}|_{\epsilon = 0} \right.  \nonumber \\& \left.
    +  P_{t,11}|_{\epsilon = 1} - P_{t,12}|_{\epsilon = 1} - P_{t,21}|_{\epsilon = 1} + P_{t,22}|_{\epsilon = 1}    \right),   \label{xt_OP}
\end{align}
where $P_{t,11}$, $P_{t,12}$, $P_{t,21}$ and $P_{t,22}$ are given in Table II, while $N$, $\tilde{D}$, $Q_3-Q_{9}$ are given in Table I.
 \begin{table}[t]
		\vspace{-0.3cm}
  \linespread{0.5}\selectfont
              	\centering
              	\caption{$N$, $\tilde{D}$, $Q_3-Q_{9}$} \vspace{-0.2cm}
				\scalebox{0.9}{	\begin{tabular}{|c|c|}
						\hline
						\textbf{Term} & \textbf{Value} \\
						\hline
						\hline
						$N$ & $(\frac{\eta}{A k_2 u_t}+\frac{\eta}{A k_2})\frac{1}{\frac{B}{A k_2 u_1}+\frac{B k_1}{A k_2}}$  \\
						\hline
						$\tilde{D}$ & $\frac{\eta u_1 C}{(1+ u_1 k_1) B u_t} (1+u_t) - \frac{\eta}{A k_2} - \frac{\eta u_2}{A}$  \\
						\hline
						$K$ & $\frac{\eta(1+\frac{1}{u_t})}{B(k_1+\frac{1}{u_1})} + \frac{\eta(u_2-\frac{1}{k_2 u_t})}{B(u_2+\frac{k_1}{k_2})}$  \\
						\hline
						$Q_3$ & $\frac{N}{\lambda_1}+\frac{B N}{A k_2 \lambda_2 u_1}  - \frac{\eta}{A k_2 \lambda_2}$   \\
						\hline
						$Q_4$ & $(\frac{1}{\lambda_1}+\frac{B}{A k_2 \lambda_2 u_1}) (\frac{\eta}{A C k_2}+\frac{\eta u_2}{A C}) - \frac{\eta}{A k_2 \lambda_2}$  \\
						\hline
						$Q_5$ & $\frac{N}{\lambda_1}+\frac{B N u_2}{A \lambda_2 } + \frac{ \eta u_2   }{A   \lambda_2 }$   \\
						\hline
						$Q_6$ & $(\frac{1}{\lambda_1}+\frac{B u_2}{A \lambda_2 \rho}) (\frac{\eta}{A C k_2}+\frac{\eta u_2}{A C}) + \frac{ \eta u_2   }{A   \lambda_2 }$  \\
						\hline
						$Q_7$ & $- (\frac{1}{\lambda_1}+\frac{B u_2}{A \lambda_2 })(\eta u_2 - \frac{\eta}{k_2 u_t})\frac{1}{\frac{B k_1}{k_2}+B u_2} + \frac{\eta u_2 }{A \lambda_2}$  \\
						\hline
						$Q_8$ & $- (\frac{1}{\lambda_1}-\frac{B k_1}{A k_2 \lambda_2})(\eta u_2 - \frac{\eta}{k_2 u_t})\frac{1}{\frac{B k_1}{k_2}+B u_2}+\frac{\eta   }{A  k_2 \lambda_2 u_t}$  \\
						\hline
						$Q_{9}$ & $\frac{N}{\lambda_1}-\frac{N B k_1}{A k_2 \lambda_2} +\frac{\eta   }{A  k_2 \lambda_2 u_t}$  \\
						\hline
				\end{tabular}} 		\label{param_values}
				\hfill
				\vspace{-0.4cm}
      \end{table} 
\begin{table*}[h] 
		\linespread{0.5}\selectfont
		\centering
		\caption{$P_{t,11}$, $P_{t,12}$, $P_{t,21}$, $P_{t,22}$}
  \vspace{-0.2cm}
        \resizebox{\textwidth}{!}{
		\begin{tabular}[t]{c|c|c}
			\hline \hline \\
		\textbf{Condition}	
        & \textbf{Term}
		& \textbf{Expression} \\  
			\hline 
			\multirow{3}{*}[-0.3cm]{\centering $\begin{aligned}
            &u_t (1+u_1 k_1)(1+k_2 u_2) \\& + k_2 u_1 u_2 < 1 
        \end{aligned}$ } & $P_{t,11}$       &   
        $\begin{aligned}- \frac{ A k_2 \lambda_2 u_1 e^{\frac{ 1}{A \rho k_2 \lambda_2 }} }{ A k_2 \lambda_2 u_1 + B \lambda_1 }   \left(  \Phi\!\left(\frac{u_2}{\tilde{D} A \rho}+\frac{1}{\tilde{D} A \rho k_2}, Q_3\right)   - e^{- (\frac{1}{\lambda_1}+\frac{B}{A k_2 \lambda_2 u_1}) (\frac{u_2}{A C \rho}+\frac{1}{A C \rho k_2})   }  \Phi\!\left(\frac{u_2}{\tilde{D} A \rho}+\frac{1}{\tilde{D} A \rho k_2},Q_4\right)    \right)\end{aligned}$  \\
         \cline{2-3}
			 
		& $P_{t,12}$ & $\begin{aligned}- \frac{ A \lambda_2  e^{-\frac{ u_2}{A  \lambda_2 \rho }} }{ A \lambda_2  + B u_2 \lambda_1 } \left(    \Phi\!\left( \frac{u_2}{\tilde{D} A \rho}+\frac{1}{\tilde{D} A \rho k_2},Q_5\right)   - e^{- (\frac{1}{\lambda_1}+\frac{B u_2}{A \lambda_2 }) (\frac{u_2}{A C \rho}+\frac{1}{A C \rho k_2})   }  
  \Phi\!\left(\frac{u_2}{\tilde{D} A \rho}+\frac{1}{\tilde{D} A \rho k_2}, Q_6 \right)    \right)\end{aligned}$ \\
			\hline  \hline
  \multirow{3}{*}{\text{otherwise}} & $P_{t,11}$
			  & $0$  \\
			 \cline{2-3}
		& $P_{t,12}$ &   $0$ \\ 
            
			\hline \hline
        \multirow{3}{*}[-0.3cm]{\centering $\begin{aligned}
            &k_2 u_2 u_1 ( 1  + u_t + k_1 u_t ) \\& + u_t (k_1 u_1 + k_2 u_2) < 1
        \end{aligned}$} & $P_{t,21}$       &   $\begin{aligned}- \frac{A \lambda_2  e^{-\frac{u_2}{A \lambda_2 \rho}}}{A \lambda_2  + B \lambda_1 u_2}  \left(  e^{ (\frac{1}{\lambda_1}+\frac{B u_2}{A \lambda_2 })(\frac{u_2}{\rho}+\frac{1}{\rho k_2})\frac{1}{\frac{B k_1}{k_2}+B u_2}} \Phi\!\left(- \frac{1}{K} \left(\frac{u_2}{A \rho}+\frac{1}{A \rho k_2}\right)\frac{1}{\frac{B k_1}{A k_2}+\frac{B u_2}{A}},Q_7\right) 
- \Phi\!\left(- \frac{1}{K} \left(\frac{u_2}{A \rho}+\frac{1}{A \rho k_2}\right)\frac{1}{\frac{B k_1}{A k_2}+\frac{B u_2}{A}},Q_8\right)  \right)\end{aligned}$  \\
         \cline{2-3} 
		& $P_{t,22}$ &   $\begin{aligned} - \frac{A k_2 \lambda_2  e^{\frac{ 1}{A \rho k_2 \lambda_2}}}{A k_2 \lambda_2  - B k_1 \lambda_1}   \left(  e^{ (\frac{1}{\lambda_1}-\frac{B k_1}{A k_2 \lambda_2})(\frac{u_2}{\rho}+\frac{1}{\rho k_2})\frac{1}{\frac{B k_1}{k_2}+B u_2}}  \Phi\!\left(- \frac{1}{K} \left(\frac{u_2}{A \rho}+\frac{1}{A \rho k_2}\right)\frac{1}{\frac{B k_1}{A k_2}+\frac{B u_2}{A}},Q_9\right) 
 -   \Phi\!\left(- \frac{1}{K} \left(\frac{u_2}{A \rho}+\frac{1}{A \rho k_2}\right)\frac{1}{\frac{B k_1}{A k_2}+\frac{B u_2}{A}},Q_{10}\right)   \right)\end{aligned}$ \\ 
			\hline  \hline 
  \multirow{3}{*}{\text{otherwise}} & $P_{t,21}$  & $0$  \\
			 \cline{2-3}
		& $P_{t,22}$ &   $0$ \\ 
			\hline \hline
		\end{tabular} }
		\label{F_i}
   \vspace{-0.6cm}
	\end{table*}
\end{thm}

\begin{IEEEproof} The proof is provided in Appendix B.

\end{IEEEproof}

Similarly with $U_1$, the OP of the BD under pSIC cannot be given via (\ref{xt_OP}). In this context, in what follows we provide a lemma that returns BD's OP under pSIC.

 \begin{lem}
 The OP of the BD under pSIC can be given by
  \begin{equation}
  P_t^{p} = 1 - \frac{1}{2} (I_t^p|_{\epsilon=0} + I_t^p|_{\epsilon=1}),
  \label{P_t,p}
  \end{equation} 
  where 
  \begin{equation}
     I_t^p = \frac{A \lambda_2 e^{-\frac{u_2}{A \lambda_2 \rho}- (\frac{1}{\lambda_1}+\frac{B u_2}{ A \lambda_2}) \frac{u_1}{B \rho}}}{A \lambda_2 + B u_2 \lambda_1} \, \Phi\! \left(\frac{u_t}{\eta \rho},Q_{1}^{p}\! \right). 
  \end{equation}
 \end{lem}

 \begin{IEEEproof}
 The OP of the BD can be given via (\ref{OP_xt_v1}) as soon as $\tilde{P}_{t}^{o}$ is appropriately updated in order to correspond to the pSIC case. Specifically, setting $k_1\!=\!k_2\!=\!0$ in (\ref{gamma_x2}), (\ref{gamma_x1}) and (\ref{gamma_xt}) and then invoking these expressions in (\ref{Pt_v1}), $\tilde{P}_{t}^{o}$ is updated as  
\begin{equation}
    \begin{split}
     \tilde{P}_{t}^{p} &= 1 - \Pr \left(   |h_2|^2 > \frac{B \rho |h_1|^2 u_2 + \eta \rho Z u_2  +  u_2  }{A \rho}, \right. \\& \left. \ \ \ \  \ \ \ \  \ \ \ \ \ \  \ \ \  |h_1|^2 > \frac{\eta \rho Z u_1 + u_1}{B \rho}, Z > \frac{u_t}{\eta \rho}   \right). 
    \end{split}   
\end{equation}
Following steps similar to those for the calculation of the probability which appears in (\ref{P1_psic_v1}), it yields
\begin{equation}
    \begin{split}
     \tilde{P}_{t}^{p} &= 1- \frac{A \lambda_2 e^{-\frac{u_2}{A \lambda_2 \rho}- (\frac{1}{\lambda_1}+\frac{B u_2}{ A \lambda_2}) \frac{u_1}{B \rho}}}{A \lambda_2 + B u_2 \lambda_1} \int_{\frac{u_t}{\eta \rho}}^{\infty}\! e^{-Q_{1}^{p} z} f_{Z}(z) \mathrm{d}z.
    \end{split} \label{P_tp}  
\end{equation}
By expressing the above integral via the assistance of the auxiliary function $\Phi(\alpha,\beta)$ defined at Lemma \ref{lem:G} and then applying (\ref{P_tp}) in (\ref{OP_xt_v1}), we get (\ref{P_t,p}). This concludes the proof.

 \end{IEEEproof}

\vspace{-0.6cm}
\subsection{Asymptotic Outage Behavior} \label{Asympt_OP}

Recall that the outage behavior of users at high SNRs offers meaningful design guidelines for practical applications. Hence, in what follows we provide a lemma that returns useful insights regarding the outage behavior of $U_1$, $U_2$ and BD at the high SNR regime.
\vspace{-0.1cm}
\begin{lem}
    In the high SNR regime, under both cases of pSIC and ipSIC, the OPs of $U_1$, $U_2$ and BD exhibit outage floors. The floor of $U_2$  
    can be obtained by (\ref{x2_OP}), while the floors of $U_1$ under ipSIC and pSIC, can be obtained by (\ref{x1_OP_v0}) and (\ref{x1_OP_pSIC}), respectively, as soon as the exponential terms of (\ref{x2_OP}), (\ref{x1_OP_v0}) and (\ref{x1_OP_pSIC}) are set equal to one. Furthermore, the outage floor of the BD, under ipSIC and pSIC, can be obtained via (\ref{xt_OP}) and (\ref{P_t,p}), respectively, as soon as the auxiliary function $\Phi(\alpha,\beta)$ is replaced with $\Phi^{\infty}(\beta)$ given in (\ref{G_asympt}) at the top of the next page and the exponential terms of the aforementioned expressions are set equal to one.
    \begin{figure*} \vspace{-0.1cm}
\begin{equation}
\Phi^{\infty}(\beta) \!\approx\! \begin{cases}
\frac{1}{\sqrt{\beta  \lambda_{tb}} (\lambda_{1t} - \lambda_{2t})} \sum\limits_{i=1}^{2} (-1)^{i+1} \sqrt{\lambda_{it}} e^{\frac{1}{2 \beta \lambda_{it} \lambda_{tb}}}  W_{-\frac{1}{2},0} \left( \frac{1}{ \beta \lambda_{it} \lambda_{tb}} \right)      , & \lambda_{1t} \ne \lambda_{2t}
\\
\frac{1}{\beta \lambda \lambda_{tb}} e^{\frac{1}{2 \beta \lambda \lambda_{tb}}} W_{-1,-\frac{1}{2}} \left( \frac{1}{ \beta \lambda_{it} \lambda_{tb}} \right)  
, & \lambda_{1t} \!=\! \lambda_{2t} \!=\! \lambda.
\end{cases}   \label{G_asympt} \vspace{-0.15cm}
\end{equation}
\vspace{-0.3cm}
\hrulefill 
\vspace{-0.4cm}
\end{figure*}
\end{lem}

\vspace{-0.1cm}
\begin{IEEEproof}
    Regarding $U_1$, $U_2$, the lemma can be proved by taking the limits of (\ref{x2_OP}), (\ref{x1_OP_v0}) and (\ref{x1_OP_pSIC}) when $\rho \!\rightarrow \!\infty$ and considering the fact that for $\tilde{\alpha},\tilde{\beta}$ real numbers, then $\lim_{\rho \rightarrow \infty} \tilde{\beta} e^{-\frac{\tilde{\alpha}}{\rho}}\! =\! \tilde{\beta}$.  
Also, regarding BD, from Table \ref{Pt_1_2}, it is obvious that for all $P_{t,11}$, $P_{t,12}$, $P_{t,21}$, $P_{t,22}$, the first arguments of function $\Phi(\alpha,\beta)$ have the form $\frac{q}{\rho}$ with $q$ being a constant value that does not depend on $\rho$; thus, as shown from (\ref{G_def}), for $\rho \!\rightarrow\! \infty$, the lower integration limit of $\Phi(\alpha,\beta)$ tends to zero. Hence, for $\rho \!\rightarrow\! \infty$, function $\Phi$ can be approximated via the first integral of (\ref{lem_v0}), i.e., by keeping only the first term of each branch in (\ref{G_apx}) as illustrated in $\Phi^{\infty}(\beta)$ provided in (\ref{G_asympt}). Hence, by replacing function $\Phi$ with $\Phi^{\infty}$ in $P_{t,11}$, $P_{t,12}$, $P_{t,21}$, $P_{t,22}$ and then taking the limits of (\ref{xt_OP}) and (\ref{P_t,p}) for $\rho\! \rightarrow \!\infty$, i.e., exploiting that $\lim_{\rho \rightarrow \infty} \tilde{\beta} e^{-\frac{\tilde{\alpha}}{\rho}} \!=\! \tilde{\beta}$, we get BD's outage~floor.    
\end{IEEEproof}
\vspace{-0.2cm}
\begin{rem}
    From the above lemma, it occurs that, under ipSIC, the outage floors of $U_1$ and BD are directly affected by the ipSIC parameters $k_1$ and $k_2$.
\end{rem}

\vspace{-0.3cm}
\section{Physical Layer Security} 
In this section the PLS of the proposed system model is investigated by extracting analytical expressions for the IPs of all $U_2$, $U_1$ and BD. Also, useful insights on how different system parameters affect network security are provided.

\vspace{-0.2cm}
\subsection{IP of $U_2$} \vspace{-0.1cm}
Given the fact that a legitimate user's or BD's message is intercepted as soon as at least one of the $M$ Eves succeeds in decoding this message, then leveraging (\ref{g_2j}), the IP of $U_2$'s message can be given as
\begin{equation}
	P^{\text{int}}_{2}=\Pr\left(\underset{j \in \{1,...,M\}}{\text{max}}\left\{\frac{ (1 - \epsilon(1-a_1)) \rho |h_{2j}|^2}{ a_2 \rho|h_{i^{*}j}|^2  + 1}\right\}>u_2^{\text{int}}\right), \label{IP_U2_v00}
\end{equation}
 where $u_2^{\text{int}}$ is the secrecy SINR threshold of $U_2$. In the following theorem, the IP of $U_2$ is provided.

 \begin{thm} \label{IP_U2}
The IP of $U_2$ can be calculated as
\begin{equation}
		P^{\text{int}}_{2}=\begin{cases}
   1 -  \frac{1}{2}\prod\limits_{j=1}^{M} \left( 1 -  \frac{\lambda_{2j}}{\lambda_{2j}+a_2\lambda_{1j}u_2^{\text{int}}}  e^{-\frac{u_2^{\text{int}}}{\lambda_{2j}\rho}} \right) \\
   -   \frac{1}{2} \prod\limits_{j=1}^{M} \left( 1-e^{-\frac{u_2^{\text{int}} }{ \lambda_{2j}(a_1 \rho - a_2 \rho u_2^{\text{int}})}} \right), & \frac{a_1}{a_2}> u_2^{\text{int}} 
   \\
   \frac{1}{2} -  \frac{1}{2}\prod\limits_{j=1}^{M} \left( 1 -  \frac{\lambda_{2j}}{\lambda_{2j}+a_2\lambda_{1j}u_2^{\text{int}}}  e^{-\frac{u_2^{\text{int}}}{\lambda_{2j}\rho}} \right),& \text{otherwise.}
	\end{cases}   \label{IP_U2_v0}
\end{equation}
 \end{thm}
\vspace{-0.2cm}
\begin{IEEEproof}
By exploiting the law of total probability, (\ref{IP_U2_v00}) can be written as
\begin{align}
	    &P^{\text{int}}_{2}=\sum_{k=1}^{2} \Pr\left(\underset{i \in \{1,2\}}{\text{rand}}{U_i}=U_k\right) \nonumber \\
	    &\times  \Pr\left(\underset{j \in \{1,...,M\}}{\text{max}}\left\{\frac{ (1 - \epsilon(1-a_1)) \rho |h_{2j}|^2}{ a_2 \rho|h_{kj}|^2  + 1}\right\}>u_2^{\text{int}}\right),
	  \label{IP_U2_v2} \normalsize
\end{align}
or equivalently 
\begin{align}
    P^{\text{int}}_{2}&=\frac{1}{2} \left\{   \Pr\left(\underset{j \in \{1,...,M\}}{\text{max}}\left\{Y^{I}_{j}\right\}>u_2^{\text{int}}\right) \right. \\& \left. \ \ \ \ \ \ \ \ \ \ \ \ \ \nonumber  + \Pr\left(\underset{j \in \{1,...,M\}}{\text{max}}\left\{Y^{II}_{j}\right\}>u_2^{\text{int}}\right)
	    \right\},
\end{align}
where
\begin{equation}
	Y^{I}_{j}=\frac{ \rho |h_{2j}|^2}{ a_2 \rho|h_{1j}|^2  + 1}, \quad Y^{II}_{j}=\frac{ a_1 \rho |h_{2j}|^2}{ a_2 \rho|h_{2j}|^2  + 1}.
\end{equation} 
Next, the CDFs of the RVs $Y^{I}_{j}$ and $Y^{II}_{j}$ will be calculated. By using the definition of a CDF, the following holds
\begin{align}
	F_{Y^{I}_{j}}(y)&\! =\!\Pr\left(Y^{I}_{j}\!<\!y\right)\! = \!\int_{0}^{\infty} \!F_{|h_{2j}|^2}\left(\!\frac{y\! +\!  a_2 \rho x y}{ \rho} \!\right)\! f_{|h_{1j}|^2}(x) \mathrm{d}x 
    \nonumber \\ 
	&= 1 -  \frac{\lambda_{2j}}{\lambda_{2j}+a_2\lambda_{1j}y}  e^{-\frac{y}{\lambda_{2j}\rho}}. 
	\label{F_Y1}
\end{align}
On the other hand, regarding RV $Y^{II}_{j}$, its CDF occurs as
\begin{equation}
	\begin{split}
		F_{Y^{II}_{j}}(y)&\!=\! \begin{cases}
			\!F_{|h_{2j}|^2} \left(\!\frac{y }{ a_1 \rho\! -\! a_2 \rho y} \!\right)\!=\!1\!-\!e^{-\frac{y }{ \lambda_{2j}(a_1 \rho \!-\! a_2 \rho y)}}       , & \frac{a_1}{a_2}> y
			\\
			\!1       ,  & \frac{a_1}{a_2} \leq y.
		\end{cases} 
	\end{split} \label{F_Y2}
\end{equation}
By exploiting (\ref{F_Y1}), (\ref{F_Y2}) and the fact that the CDF of RV $\tilde{Y}\!=\!\underset{j \in \{1,...,M\}}{\text{max}}\left\{Y_{j}\right\}$, if $Y_j$ are independent with each other, can be given as 
\begin{equation} 
F_{\tilde{Y}}(y)= \prod_{j=1}^{M} F_{Y_j}(y),  \label{CDF_MAX}
\end{equation} 
then (\ref{IP_U2_v2}) can be transformed into
\begin{equation}
	\begin{split}
		P^{\text{int}}_{2}=\frac{1}{2} \left\{ 2 -  \prod_{j=1}^{M} F_{Y^I_j}(u_2^{\text{int}})  -   \prod_{j=1}^{M} F_{Y^{II}_j}(u_2^{\text{int}}) \right\}.
	\end{split}   \label{IP_U2_v3}
\end{equation}
This concludes the proof.

\end{IEEEproof}



\vspace{-0.2cm}
\begin{rem} \label{rem_U2}
From (\ref{IP_U2_v0}), it occurs that as $a_1$ increases, i.e., as the power utilized for the transmission of the friendly artificial noise reduces, the IP of $U_2$ also increases.     
\end{rem}

\vspace{-0.3cm}
\subsection{IP of $U_1$}
The IP of $U_1$ can be extracted in a similar manner with the IP of $U_2$ and is provided in the following theorem.
 \begin{thm}
The IP of $U_1$ can be evaluated as
\begin{equation}
	P^{\text{int}}_{1}=\begin{cases}
   1 -  \frac{1}{2}\prod\limits_{j=1}^{M} \left( 1 -  \frac{\lambda_{1j}}{\lambda_{1j}+a_2\lambda_{2j}u_1^{\text{int}}}  e^{-\frac{u_1^{\text{int}}}{\lambda_{1j}\rho}} \right) \\
   -   \frac{1}{2} \prod\limits_{j=1}^{M} \left( 1-e^{-\frac{u_1^{\text{int}} }{ \lambda_{1j}(a_1 \rho - a_2 \rho u_1^{\text{int}})}} \right), & \frac{a_1}{a_2}> u_1^{\text{int}} 
   \\
   \frac{1}{2} -  \frac{1}{2}\prod\limits_{j=1}^{M} \left( 1 -  \frac{\lambda_{1j}}{\lambda_{1j}+a_2\lambda_{2j}u_1^{\text{int}}}  e^{-\frac{u_1^{\text{int}}}{\lambda_{1j}\rho}} \right),& \text{otherwise},
	\end{cases}   \label{IP_U1_v0}
\end{equation}
where $u_1^{\text{int}}$ is the secrecy SINR threshold of $U_1$.
 \end{thm}

 \begin{IEEEproof}
     The proof can be straightforwardly obtained by following similar lines with the proof of Theorem \ref{IP_U2}. 
 \end{IEEEproof}

 At this point, it is noted that, similarly with Remark \ref{rem_U2}, as $a_1$ increases, the IP of $U_1$ also increases. 

\vspace{-0.3cm}
\subsection{IP of BD}
Taking into account (\ref{g_tj}), BD's IP can be expressed as 
\begin{equation}
	P^{\text{int}}_{t}=\Pr\left(\underset{j \in \{1,...,M\}}{\text{max}}\left\{\frac{ \eta \rho |h_{tj}|^2 (|h_{2t}|^2+|h_{1t}|^2)}{ a_2 \rho|h_{i^{*}j}|^2  + 1}\right\}>u_t^{\text{int}}\right)\!,
	\label{IP_Ut}
\end{equation}
where $u_t^{\text{int}}$ is the secrecy SINR threshold for the BD. In the following theorem, BD's IP is obtained.

\begin{thm}  \label{thm:IP_BD}
BD's IP can be evaluated as
\begin{equation}
	\begin{split}
		P^{\text{int}}_{t}&= 1 - \frac{1}{2}  \sum\limits_{k=1}^{2} I_{kj},
	\end{split}   \label{IP_Ut_vv0}
\end{equation}
where $I_{kj}$ is provided via (\ref{Ikj_f}) at the top of the next page with $x_n$ being the $n$-th root of Laguerre polynomial $L_{\tilde{\Delta}}(x)$, $w_n$ being weights given from $w_n=\frac{x_n}{(\tilde{\Delta}+1)^2 (L_{\tilde{\Delta}+1}(x_n))^2}$ \cite{abramowitz1968handbook} and $\tilde{\Delta}$ denoting an accuracy-complexity trade-off parameter.
\end{thm}

\begin{figure*} \vspace{-0.4cm}
\begin{equation}
	I_{kj}\approx \begin{cases}
		\frac{1}{\lambda_{1t}-\lambda_{2t}} \sum\limits_{i=1}^{2} (-1)^{i+1} \lambda_{it}   \sum\limits_{n=1}^{\tilde{\Delta}} w_n \prod_{j=1}^{M} \left( 1 - \frac{\eta \lambda_{tj} e^{-\frac{u_t^{\text{int}}}{\eta \rho \lambda_{tj} \lambda_{it} x_n}}}{\eta \lambda_{tj} + a_2 \lambda_{kj} \frac{u_t^{\text{int}}}{\lambda_{it} x_n}} \right)  , & \lambda_{1t} \ne \lambda_{2t}
		\\
		 \sum\limits_{n=1}^{\tilde{\Delta}} w_n \,  x_n \prod_{j=1}^{M} \left( 1 - \frac{\eta \lambda_{tj} e^{-\frac{u_t^{\text{int}}}{\eta \rho \lambda_{tj} \lambda x_n}}}{\eta \lambda_{tj} + a_2 \lambda_{kj} \frac{u_t^{\text{int}}}{\lambda x_n}} \right)   , & \lambda_{1t} = \lambda_{2t} = \lambda
	\end{cases}  \label{Ikj_f} \vspace{-0.1cm}
\end{equation}
\hrulefill 
\vspace{-0.7cm}
\end{figure*} 

\begin{IEEEproof} The proof is provided in Appendix C.
 
\end{IEEEproof}

\vspace{-0.2cm}
\begin{rem} \label{IP_a1_BD} 
    From (\ref{IP_Ut_vv0}) it holds that, similarly with the cases of $U_1$'s and $U_2$'s IPs, as $a_1$ increases, BD's IP also increases. 
\end{rem}

\vspace{-0.5cm}
\subsection{High SNR regime}

Given the fact that in order to enhance the system model's PLS, we devote a portion of users' transmit power for the transmission of an artificial noise, it is considered important to investigate the IPs of $U_1$, $U_2$ and BD at high SNRs. In this direction, in what follows we present a lemma that provides useful insights regarding the IPs of $U_1$, $U_2$ and BD in the high SNR regime.

\begin{lem}
    In the high SNR regime, the IPs of $U_1$, $U_2$ and BD reach constant values that can be obtained by (\ref{IP_U2_v0}), (\ref{IP_U1_v0}) and (\ref{IP_Ut_v0}) as soon as the exponential terms of the aforementioned expressions are set equal to one.
\end{lem}
\vspace{-0.1cm}
\begin{IEEEproof}
    The lemma can be proved by taking the limits of the expressions (\ref{IP_U2_v0}), (\ref{IP_U1_v0}) and (\ref{IP_Ut_v0}) when $\rho \rightarrow \infty$ and considering the fact that for $\tilde{\alpha},\tilde{\beta}$ real numbers, then $\lim_{\rho \rightarrow \infty} \tilde{\beta} e^{\frac{\tilde{\alpha}}{\rho}} = \tilde{\beta}$.  
\end{IEEEproof}

\vspace{-0.1cm}
\begin{rem} \label{IP_floors}
    From the above lemma, it occurs that the asymptotic values that the IPs of $U_1$, $U_2$ and BD reach at high SNRs are directly affected by the power allocation coefficients $a_1$, $a_2$, i.e., by the portion of the total users' power utilized for the transmission of the artificial noise.
\end{rem}

\vspace{-0.3cm}
\section{Numerical Results and Discussion}
\vspace{-0.1cm}

In this section, both simulations (sim.) and analytical (ana.)
results are presented in order to validate the provided analysis
and to evaluate the system model’s outage performance and PLS versus various system parameters under both pSIC and ipSIC cases. For the extraction of the following figures, unless otherwise stated, $\lambda_{1}\!=\!0.1$, $\lambda_2\!=\!1.5$, $\lambda_{1t}\!=\!\lambda_{tb}\!=\!0.4$, $\lambda_{2t}\!=\!0.5$ are assumed. Furthermore, $a_1\!=\!0.8$, $R_1\!=\!R_2\!=\!0.5$, $R_t\!=\!0.05$ bps/Hz as well as $\eta\!=\!0.01$ are considered. Also, $M\!=\!3$, $\lambda_{ij}\!=\!0.15$, $\lambda_{tj}\!=\!0.1$ and $u_1^{\text{int}}\!=\!0.4$, $u_2^{\text{int}}\!=\!0.3$, $u_t^{\text{int}}\!=\!0.03$ are adopted. Finally, for simplicity, we assume $k_1\!=\!k_2\!=\!k$. 

\begin{figure} 	
\vspace{-0.05cm}
	\centering	\includegraphics[keepaspectratio,width=0.75\columnwidth]{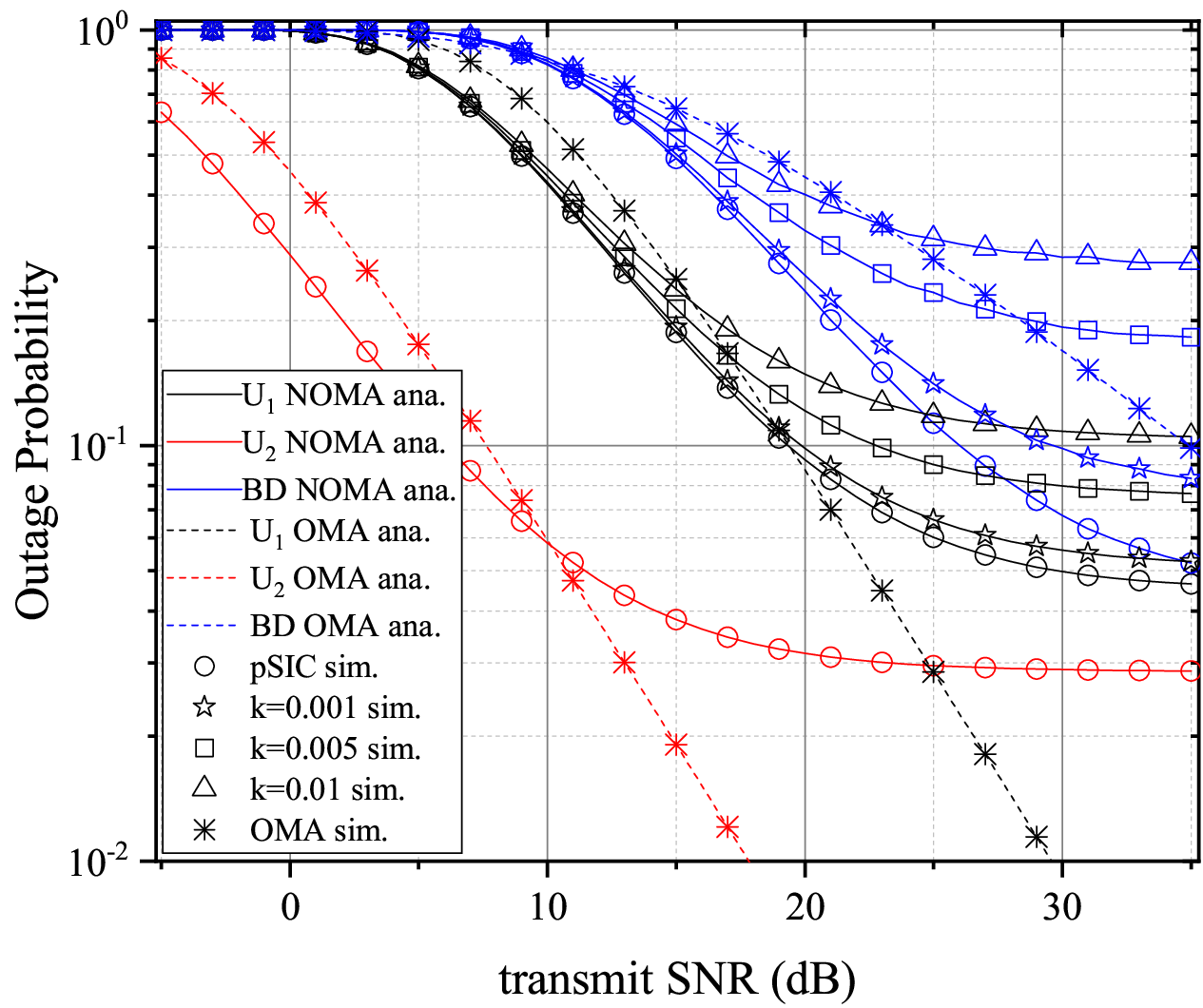}
	\caption{NOMA-users' $U_1$, $U_2$ and BD's OPs versus $\rho$.}    \label{Fig:OP_vs_rho} 
   \vspace{-0.5cm}
\end{figure}

In Fig. \ref{Fig:OP_vs_rho}, NOMA-users' as well as BD's OP are depicted versus transmit SNR $\rho$. Moreover, an OMA-AmBC scheme is provided as a benchmark. 
It can be shown that analytical results coincide with simulation ones, thus the authenticity of the presented analysis is verified. Regarding $U_2$, it is illustrated that its OP is not affected via the ipSIC since its message is decoded first at the BS side as well as that the proposed scheme outperforms OMA until $\rho$ reaches 10 dBs and then OMA achieves lower OPs. On the other hand, in terms of $U_1$, NOMA outperforms OMA for the whole region of $\rho\! \in\! [-5,18]$ dBs when pSIC or ipSIC with $k\! =\! 0.001$ is applied. It is also interesting to observe that even for the case of ipSIC with $k \!=\! 0.01$, NOMA outperforms OMA at the low-to-medium SNR region.
In terms of the BD, it can be observed that, under both pSIC
or ipSIC case with $k \!=\! 0.001$, the proposed scheme
outperforms OMA for the whole SNR region. Finally, it is
shown that for increased $k$ values, the OPs of both $U_1$ and BD increase as well as that all $U_1$, $U_2$ and BD reach outage floors at high SNRs which are directly affected by the ipSIC parameters $k_1$, $k_2$; thus, the analysis presented in section \ref{Asympt_OP} is validated. 

\begin{figure} 	
	\centering	\includegraphics[keepaspectratio,width=0.75\columnwidth]{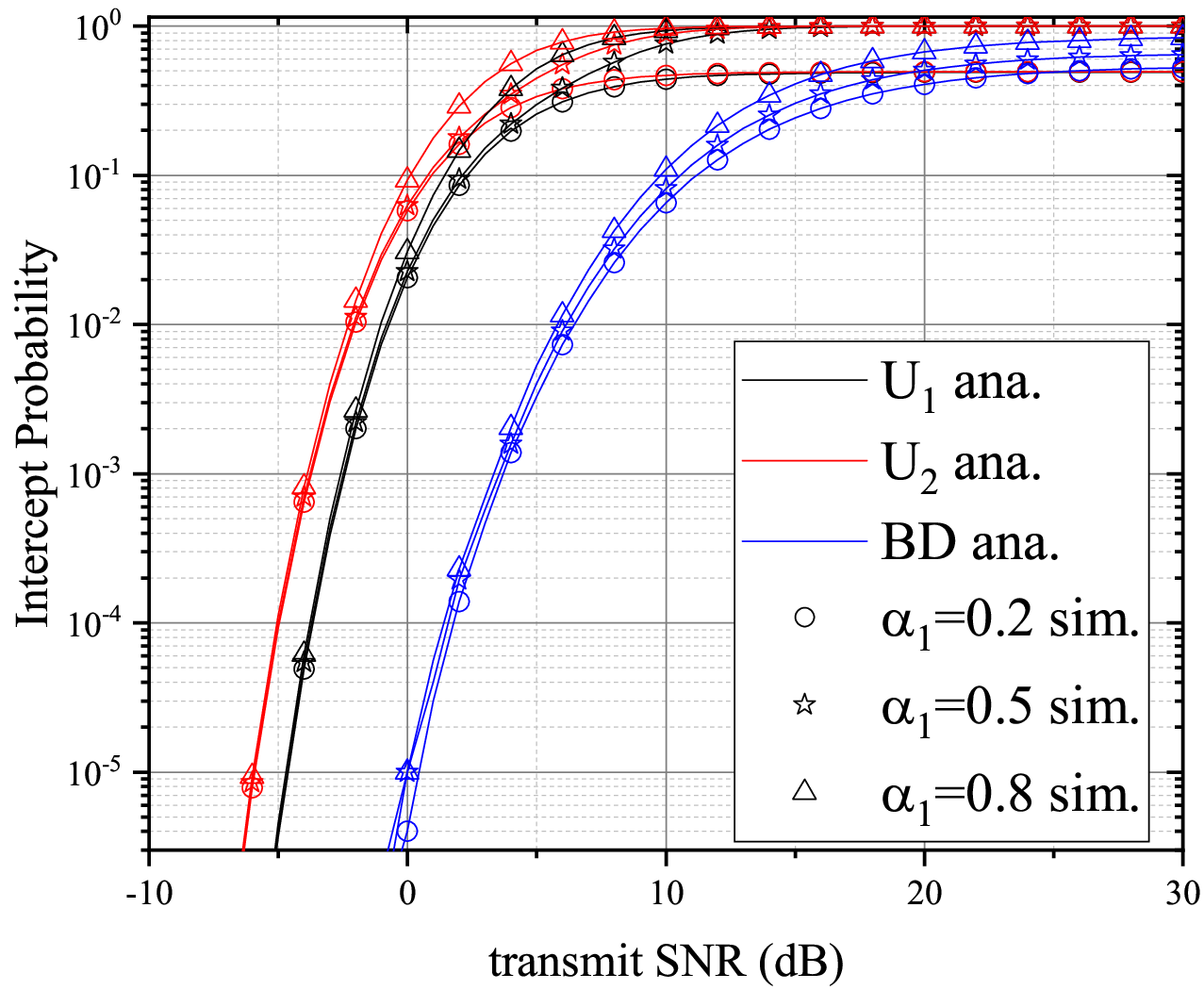}
	\caption{NOMA-users' $U_1$, $U_2$ and BD's IPs versus $\rho$.}    \label{Fig:IP_vs_rho} 
   \vspace{-0.5cm}
\end{figure} 

In Fig. \ref{Fig:IP_vs_rho}, NOMA-users' and BD's IPs are presented along with transmit SNR $\rho$ for various power allocation factor $a_1$ values when $\lambda_{ij}\!=\!\lambda_{tj}\!=\!0.1$. For fixed $a_1$, as pointed out in Remarks \ref{rem_U2} and \ref{IP_a1_BD}, increasing $\rho$ leads to increased IPs for all nodes. Furthermore, it can be observed that IP floors appear at high SNRs that are directly affected from the $a_1$ values which was also revealed by Remark \ref{IP_floors}. This is an important observation since with a careful selection of parameter $a_1$, system's security behavior can be improved. Furthermore, it can be observed that increased $a_1$ leads to worse IP behavior for all nodes. 
Taking into consideration the observations from Fig. 1, an interesting trade-off arises according to which while in terms of OP, increased $\rho$ is desired, this also leads to increased likelihood of eavesdropping. Hence, when designing NOMA-enabled AmBC networks, users' transmission power must be carefully selected and optimized. 

\begin{figure*}[t]
\vspace{-0.2cm}
\centering
\begin{minipage}[t]{.33\textwidth}
\centering
	\includegraphics[keepaspectratio,width=0.99\columnwidth]{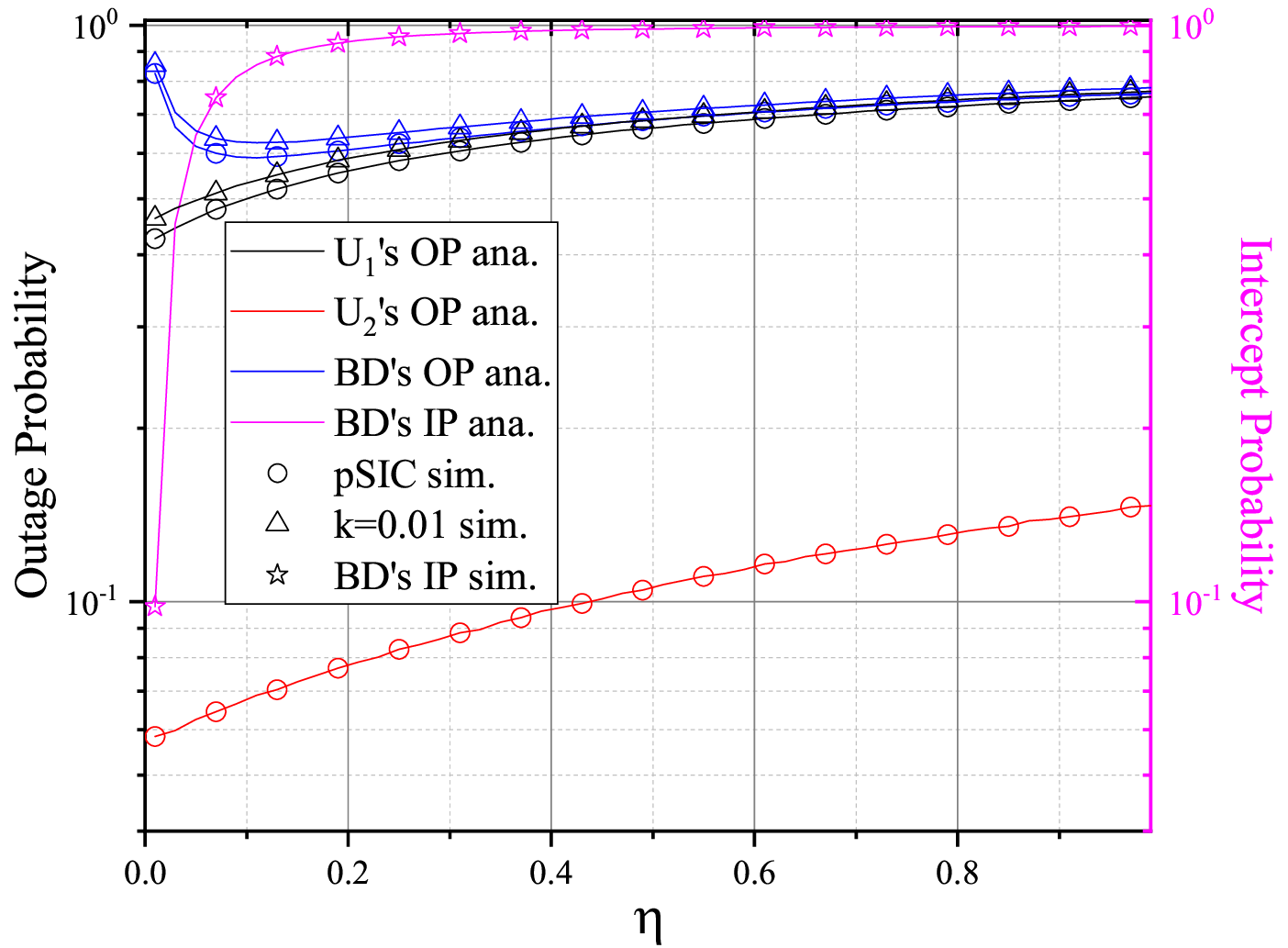}
  \vspace{-0.45cm} 
 \captionsetup{margin=0.3cm}
 \caption{NOMA-users' $U_1$, $U_2$ OPs and BD's OP and IP versus $\eta$.}
	\label{Fig:IP_vs_eta}
 \end{minipage}%
\begin{minipage}[t]{.33\textwidth}	\centering
	\centering	\includegraphics[keepaspectratio,width=0.88\columnwidth]{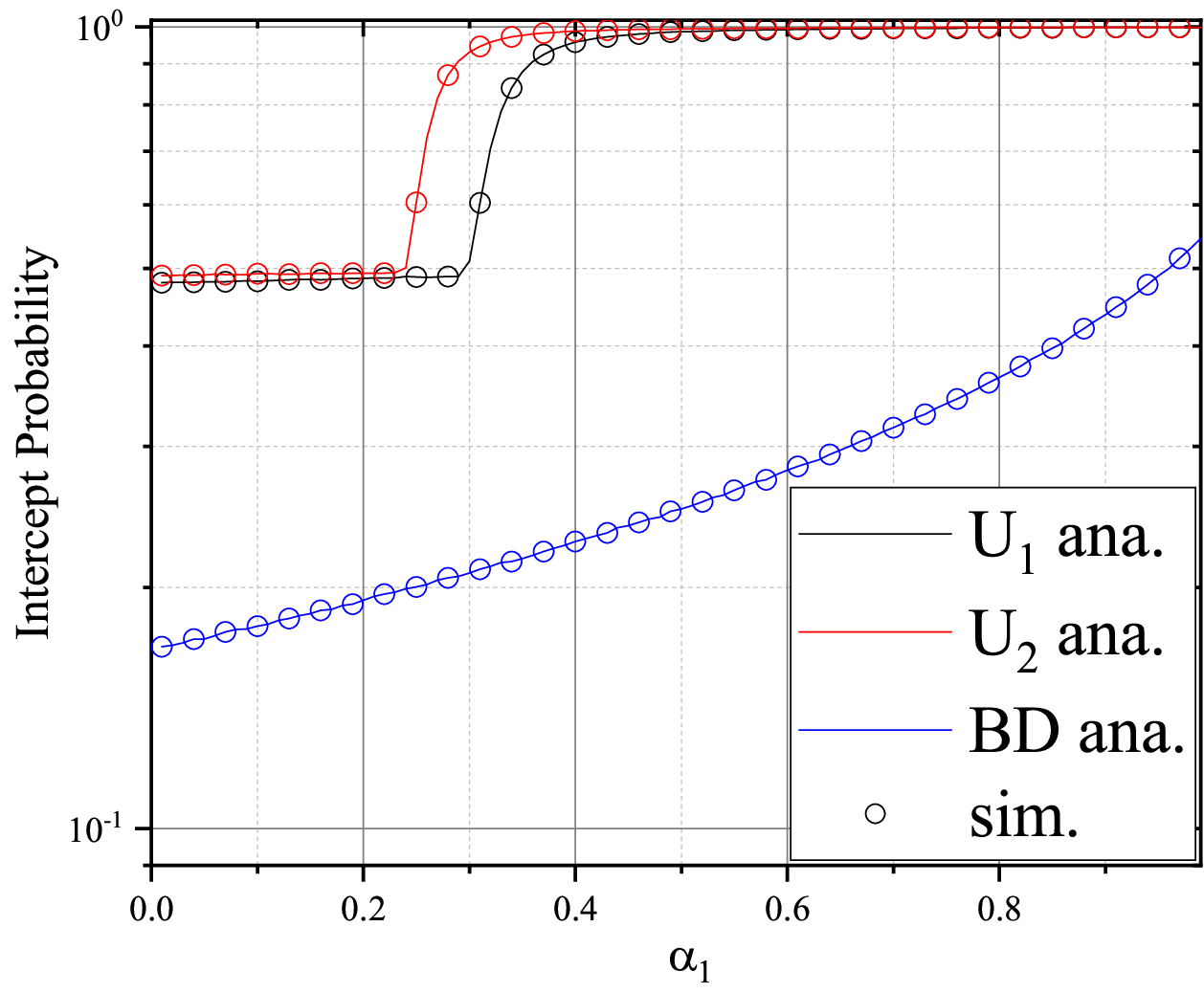}
  \vspace{-0.45cm}  
 \captionsetup{margin=0.3cm}
  \vspace{+0.4cm}
 \caption{NOMA-users' $U_1$, $U_2$ and BD's IPs versus $a_1$.}
	\label{Fig:IP_vs_a1}
 \end{minipage}%
\begin{minipage}[t]{.33\textwidth}
	\centering
	\centering	\includegraphics[keepaspectratio,width=0.90\columnwidth]{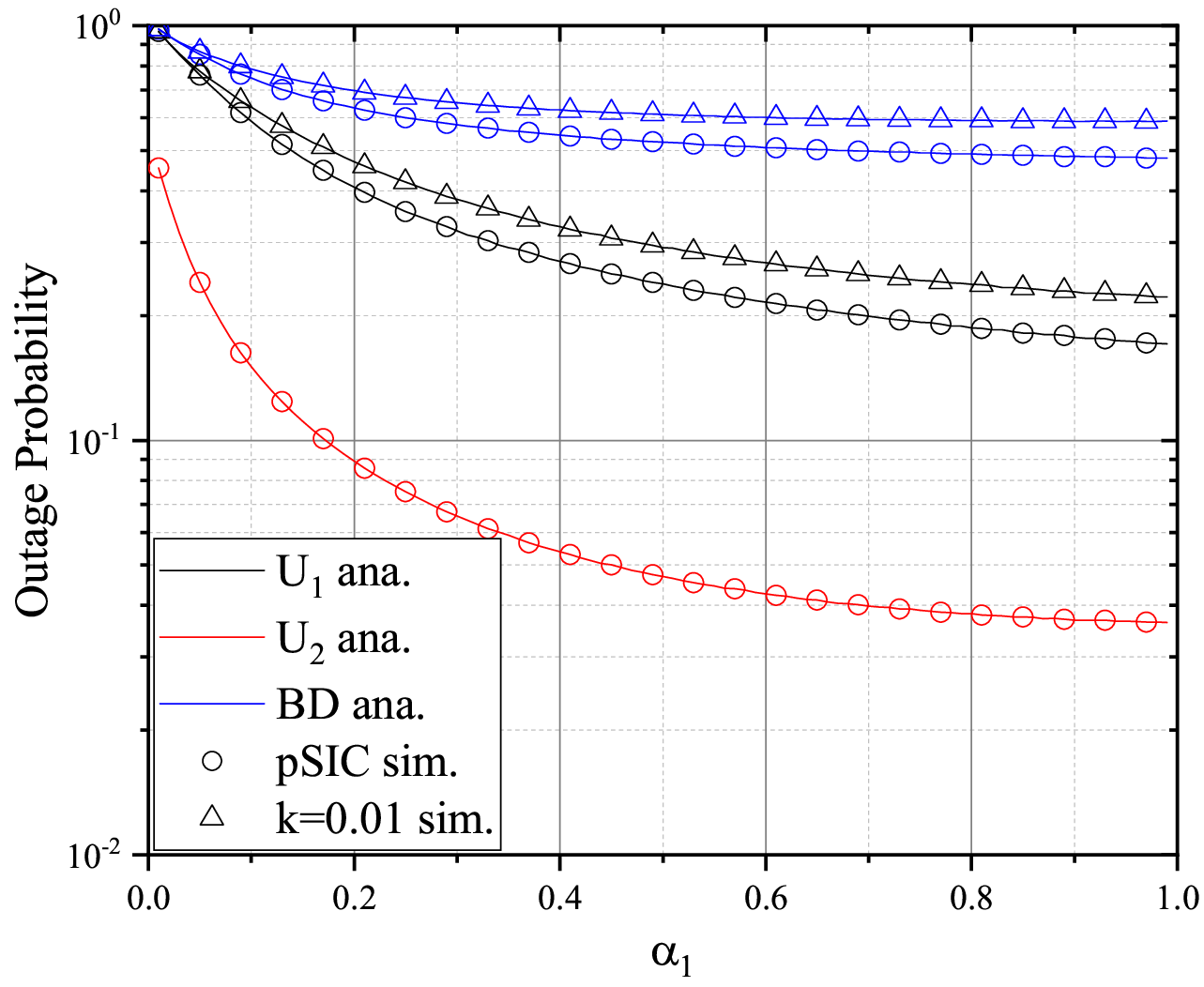}
  \vspace{-0.45cm} 
 \captionsetup{margin=0.3cm}
 \vspace{+0.4cm}
 \caption{NOMA-users' $U_1$, $U_2$ and BD's OPs versus $a_1$.}
	\label{Fig:OP_vs_a1}
 \end{minipage}
 \vspace{-0.1in}
\vspace{-0.3cm}
\end{figure*} 

In Fig. \ref{Fig:IP_vs_eta}, users $U_1$ and $U_2$ as well as BD's OPs with respect to the reflection coefficient $\eta$ are depicted when $\rho\!=\!10$ dBs. It can be seen that as $\eta$ increases, U1’s and U2’s OPs also
increase. This happens since, as (\ref{gamma_x2}) and (\ref{gamma_x1}) witness, increased
$\eta$ leads to increased interference during the decoding of $x_1$ and $x_2$. On the other hand, BD’s outage performance versus $\eta$ shows a convex behavior, i.e., BS’s OP decreases up to a specific $\eta$ value, reaches a curve point and then begins to increase. This may seem contradictory since as $\eta$ increases, the SINR for successful decoding of $x_t$ given in (\ref{gamma_xt}) also increases. However,
as (\ref{Pt_v1}) indicates, for the successful decoding of $x_t$ it is also
necessary that $x_2$ and $x_1$ are also successfully decoded. Hence,
the fact that increased $\eta$ leads to increased $\gamma_{x_t}$ but decreased $\gamma_{x_1}$ and $\gamma_{x_2}$ explains the aforementioned convex behavior. From the above, it occurs that there exist an interesting trade-off regarding NOMA-users' and BD's OPs. More specifically, NOMA-users require as low $\eta$ as possible, while BD desires to operate under the $\eta$ value which enables its OP to be minimized, i.e., the $\eta$ value under which the aforementioned curve point is achieved. Moreover, in Fig. \ref{Fig:IP_vs_eta}, BD's IP is presented alongside $\eta$ with magenta color. It can be shown that as $\eta$ increases, BD's IP also increases. This is expected since as (\ref{g_tj}) clearly depicts, increased $\eta$ leads to increased SINR for successful eavesdropping at the Eves side, and thus, it is more likely that Eves intercept BD's message. Hence, the moderately small $\eta$ values met in practical applications are beneficial when it comes to BD's security. 

In Fig. \ref{Fig:IP_vs_a1}, NOMA-users' and BD's IPs versus power allocation coefficient $a_1$ are presented when $\rho\!=\!15$ dBs. It is obvious that increased $a_1$ leads to increased BD's IP. This happens since as $a_1$ increases, then $a_2$ decreases, and thus, the amount of the total power used to emit the artificial noise also decreases; under these conditions, Eves are more likely to intercept BD's message $x_t$. This was also revealed from Remark \ref{IP_a1_BD}. Furthermore, as also SINR expressions (\ref{g_2j}) and (\ref{g_1j}) as well as Remark \ref{rem_U2} witness, increased $a_1$ leads to increased IPs for messages $x_1$ and $x_2$. Specifically, for both $U_1$ and $U_2$, it can be observed that as $a_1$ increases, their IPs show an initial robust behavior, but there exist an $a_1$ value, called $a_1^{*}$, where the further increase of $a_1$ leads to a rapid increase of $U_1$'s and $U_2$'s IPs. This can be explained by observing (\ref{IP_U2_v0}) and (\ref{IP_U1_v0}). Specifically, when $a_1\!<\!a_1^{*}$ then the condition $\frac{a_1}{a_2}\!>\! u_i^{\text{int}}$, with $i \in \{1,2\}$, does not hold; thus, NOMA-users' IPs remain relatively constant. On the other hand, a further increase of $a_1$, i.e., $a_1\!>\!a_1^{*}$, allows the condition $\frac{a_1}{a_2}\!>\!u_i^{\text{int}}$ to hold, and as (\ref{IP_U2_v0}) and (\ref{IP_U1_v0}) indicate, then another term is generated which contributes to the maximization of NOMA-users' IPs. 

In Fig. \ref{Fig:OP_vs_a1}, NOMA-users' and BD's OPs are depicted along with power allocation factor $a_1$ when $\rho\!=\!15$ dBs. As expected, increased $k$ leads to increased $U_1$'s and BD's OPs. Furthermore, it can be observed that as $a_1$ increases, $U_1$'s, $U_2$'s as well as BD's OPs decrease. 
Hence, taking also into account the observations from Fig. \ref{Fig:IP_vs_a1}, an interesting trade-off arises. Specifically, as $a_1$ increases, then NOMA-users' and BD's OPs decrease, but in the meantime, all messages are more susceptible to eavesdropping.  

\vspace{-0.25cm}
\section{Conclusion} \vspace{-0.1cm}
In this study, a novel analytical framework for studying the reliability and PLS of uplink NOMA-enabed AmBC systems was introduced. Specifically, assuming a network consisting of multiple legitimate users, passive IoT BDs and Eves, we derived novel closed-form expressions for NOMA-users' and BD's OPs for both pSIC and ipSIC. Moreover, assuming that a legitimate user is selected to act as a friendly jammer, we investigated the PLS of the proposed system by extracting NOMA-users' and BD's IPs. To gain more insights, an asymptotic analysis for the high SNR regime was conducted, revealing the appearance of outage floors for NOMA-users and BDs in the high SNR regime under both pSIC and ipSIC cases, as well as that NOMA-users' and BD's IPs tend to reach constant values at high SNRs. Finally, the validity of the provided theoretical results was carried out via simulation results which also compared the performance of the proposed NOMA-based AmBC approach with a conventional OMA-based scheme, illustrated the impact of different system parameters on NOMA-users' and BDs' outage performance and IPs as well as revealed interesting trade-offs regarding system's reliability and security. 

	

 \vspace{-0.25cm}
\section*{Appendix A: Proof of Lemma \ref{lem:G}} \vspace{-0.1cm}
By utilizing the unit step function, defined as 
\begin{equation}
    U(x) = \begin{cases}
        1 , & x>0 \\
        0, & x<0,
    \end{cases}
\end{equation}
then (\ref{G_def}) can be rewritten as
    \begin{equation}
       \Phi(\alpha, \beta) = \int_{0}^{\infty} U\left(\frac{z}{a}-1\right) e^{- \beta z} f_{Z}(z) \mathrm{d}z.  \label{G_v2}
    \end{equation}
Focusing on the case when $\lambda_{1t} \! \neq \! \lambda_{2t}$, by substituting the first branch of (\ref{F_Z_v6}) in (\ref{G_v2}) and then expressing the unit step, the exponential and the Bessel functions appearing in the resultant expression via equivalent Meijer-G function representations \cite[Chap. 8.4]{prudnikov1986integrals}, we get
    \begin{align}
       \Phi(\alpha, \beta) &= \frac{2}{\lambda_{tb}(\lambda_{1t}-\lambda_{2t})} \sum_{i=1}^{2} (-1)^{i+1} \! \int_{0}^{\infty} \MeijerG*{0}{1}{1}{1}{1}{0}{\frac{z}{\alpha}}  \nonumber \\
       & \times \MeijerG*{1}{0}{0}{1}{-}{0}{\beta z} \, \MeijerG*{2}{0}{0}{2}{-,-}{0,0}{\frac{z}{\lambda_{it} \lambda_{tb}}} \mathrm{d}z, 
  \label{G_v12}
   \end{align} 
where $G_{p,q}^{m,n}(\cdot)$ denotes the Meijer-G function. At this point, exploiting the relation between Meijer-G function and Fox-H function \cite[(6.2.8)]{springer1979algebra}, it yields
    \begin{align}
      & \Phi(\alpha, \beta) = \frac{2}{\lambda_{tb}(\lambda_{1t}-\lambda_{2t})} \sum_{i=1}^{2} (-1)^{i+1} \! \int_{0}^{\infty} \FoxH*{0}{1}{1}{1}{(1,1)}{(0,1)}{\frac{z}{\alpha}} \nonumber \\
       & \times\FoxH*{1}{0}{0}{1}{-}{(0,1)}{\beta z} \, \FoxH*{2}{0}{0}{2}{-,-}{(0,1),(0,1)}{\frac{z}{\lambda_{it} \lambda_{tb}}} \mathrm{d}z,  
  \label{G_v22}
   \end{align}
where $H_{p,q}^{m,n}(\cdot)$ denotes the Fox-H function. Finally, leveraging \cite[(2.3)]{mittal1972integral}, the first branch of (\ref{G_ex}) is obtained.

On the other hand, when $\lambda_{1t}\!=\!\lambda_{2t}$, then by invoking the second branch of (\ref{F_Z_v6}) in (\ref{G_v2}) and expressing the appearing functions via the Meijer-G function, (\ref{G_v2}) becomes
\begin{align}
       \Phi(\alpha, \beta) = \frac{2}{\lambda \lambda_{tb} \sqrt{\lambda \lambda_{tb}}}  & \int_{0}^{\infty} \sqrt{z} \, \MeijerG*{0}{1}{1}{1}{1}{0}{\frac{z}{\alpha}} \MeijerG*{1}{0}{0}{1}{-}{0}{\beta z} \nonumber \\ &  \times 
       \MeijerG*{2}{0}{0}{2}{-,-}{\frac{1}{2},-\frac{1}{2}}{\frac{z}{\lambda \lambda_{tb}}} \mathrm{d}z.
  \label{G_v3}
\end{align}
By exploiting \cite[(6.2.8)]{springer1979algebra} for expressing the Meijer-G functions via the equivalent Fox-H functions and then utilizing \cite[(2.3)]{mittal1972integral}, we get the second branch of (\ref{G_ex}).

On the other hand, towards the extraction of a useful approximation for $\Phi(\alpha,\beta)$, starting from (\ref{G_def}), it yields
      \begin{equation}
      \begin{split}
        \int_{\alpha}^{\infty} e^{- \beta z} f_{Z}(z) \mathrm{d}z &= \int_{0}^{\infty} e^{- \beta z} f_{Z}(z) \mathrm{d}z \!-\! \int_{0}^{\alpha} e^{- \beta z} f_{Z}(z) \mathrm{d}z.
        \end{split}  \label{lem_v0}
    \end{equation}  
At this point, we note that the first integral of (\ref{lem_v0}) can be calculated in a closed-form by substituting $f_{Z}(z)$ given in (\ref{F_Z_v6}) and utilizing \cite[(6.614.4) and (6.643.6)]{grad}. In the meantime, inconveniently, when invoking $f_{Z}(z)$, it occurs that the second integral of (\ref{lem_v0}) cannot be extracted in a closed-form. Nevertheless, it can be approximated via the well-known Gauss-Chebyshev quadrature \cite{hildebrand1987introduction}. From the above observations, (\ref{G_apx}) can be easily derived.  

\vspace{-0.25cm}
\section*{Appendix B: proof of theorem \ref{thm:OP_BD}}
Applying (\ref{gamma_x2}), (\ref{gamma_x1}) and (\ref{gamma_xt}) in (\ref{Pt_v1}) and utilizing the law of total probability, BD's OP can be expressed as
 \begin{equation}
	\begin{split}
		P_{t}^{o} &\!=\! \Pr\left(\!\underset{i \in \{1,2\}}{\text{rand}}{U_i}\!=\!U_1\!\right)\! \tilde{P}_{t}^{o}|_{\epsilon \!=\! 0} \!+\! \Pr\left(\!\underset{i \in \{1,2\}}{\text{rand}}{U_i}\!=\!U_2\!\right) \! \tilde{P}_{t}^{o}|_{\epsilon\! =\! 1},   \label{OP_xt_v1} 
	\end{split}	
\end{equation}
where $\tilde{P}_{t}^{o}$ can be given as
\begin{equation}
    \begin{split}
     &\tilde{P}_{t}^{o} = 1 - \Pr \left(   \frac{B \rho |h_1|^2 u_2 + \eta \rho Z u_2  +  u_2  }{A \rho} < |h_2|^2 <  \right. \\& \left. \min \!\Bigg\{ \!\frac{B \rho |h_1|^2 - \eta \rho Z u_1 - u_1}{A \rho k_2 u_1} ,  \frac{\eta \rho Z - B k_1 \rho |h_1|^2 u_t - u_t}{A \rho k_2 u_t}\! \Bigg\} \! \right)\!.
    \end{split}  \label{Pt_v2}  
\end{equation}
From (\ref{Pt_v2}), it can be observed that in order to proceed into the calculation of $\tilde{P}_{t}^{o}$, we must determine for which $|h_1|^2$, $Z$ values the following inequalities hold
\begin{equation}
    \frac{B \rho |h_1|^2 - \eta \rho Z u_1 - u_1}{A \rho k_2 u_1} < \frac{\eta \rho Z - B k_1 \rho |h_1|^2 u_t - u_t}{A \rho k_2 u_t}, \label{Pt_v3}
\end{equation}
\begin{equation}
    \frac{B \rho |h_1|^2 - \eta \rho Z u_1 - u_1}{A \rho k_2 u_1} > \frac{\eta \rho Z - B k_1 \rho |h_1|^2 u_t - u_t}{A \rho k_2 u_t}. \label{Pt_v4}
\end{equation}
After some algebraic manipulations, (\ref{Pt_v3}) can be rewritten as
\begin{equation}
    |h_1|^2 \left(\frac{B}{A k_2 u_1}+\frac{B k_1}{A k_2}\right) < Z \left(\frac{\eta}{A k_2 u_t}+\frac{\eta}{A k_2}\right).
    \label{Pt_v5}
\end{equation}
Taking into account (\ref{Pt_v5}), (\ref{Pt_v2}) can be written as
\begin{equation}
    \begin{split}
     \tilde{P}_{t}^{o} &= 1 - P_{t,1} - P_{t,2}, 
    \end{split}  \label{Pt_v6}  
\end{equation}
where
\begin{equation}
    \begin{split}
     &P_{t,1} \!=\!  \Pr \left( \!|h_1|^2 (\frac{B}{A k_2 u_1}+\frac{B k_1}{A k_2}) < Z (\frac{\eta}{A k_2 u_t}+\frac{\eta}{A k_2}), \right. \\& \left. \frac{B \rho |h_1|^2 u_2 + \eta \rho Z u_2  +  u_2  }{A \rho} \!<\! |h_2|^2 \!<\!   \frac{B \rho |h_1|^2 \!-\! \eta \rho Z u_1 \!-\! u_1}{A \rho k_2 u_1} \!\right)\!,
    \end{split}  \label{Pt_v7}  
\end{equation}
\begin{equation}
    \begin{split}
     &P_{t,2} \!=\!  \Pr \left( \!|h_1|^2  > Z (\frac{\eta}{A k_2 u_t}+\frac{\eta}{A k_2}) \frac{1}{(\frac{B}{A k_2 u_1}+\frac{B k_1}{A k_2})},  \right. \\& \left. \!\frac{B \rho |h_1|^2 u_2 \!+\! \eta \rho Z u_2  \!+\!  u_2  }{A \rho} \!<\! |h_2|^2 \!<\! \frac{\eta \rho Z \!-\! B k_1 \rho |h_1|^2 u_t \!-\! u_t}{A \rho k_2 u_t}\!\right)\!.
    \end{split}  \label{Pt_v8}  
\end{equation}
In what follows, we focus on the calculation of $P_{t,1}$ and $P_{t,2}$. Starting from (\ref{Pt_v7}), as it was analytically shown in the proof of Theorem \ref{thm_U1}, the second term of the probability holds only when $C>0$ and $|h_1|^2 > D_Z$, where $D_Z$ can be obtained by substituting $z$ with $Z$ in the expression of $D_z$ given in (\ref{P1_v5}). Hence, (\ref{Pt_v7}) becomes
\begin{equation}
    \begin{split}
     &P_{t,1} =  \Pr \left( \! D_Z < |h_1|^2 <
       Z (\frac{\eta}{A k_2 u_t}+\frac{\eta}{A k_2}) \frac{1}{(\frac{B}{A k_2 u_1}+\frac{B k_1}{A k_2})}, \right. \\& \left.  
     \!\frac{B \rho |h_1|^2 u_2 + \eta \rho Z u_2  +  u_2  }{A \rho} \!<\! |h_2|^2 \!<\! \frac{B \rho |h_1|^2 \!-\! \eta \rho Z u_1 \!-\! u_1}{A \rho k_2 u_1}\!\right)\!.
    \end{split}  \label{Pt_v9}  
\end{equation}
For (\ref{Pt_v9}) to take non-zero values, the following must hold
\begin{equation}
   D_Z <  Z (\frac{\eta}{A k_2 u_t}+\frac{\eta}{A k_2})\frac{1}{\frac{B}{A k_2 u_1}+\frac{B k_1}{A k_2}},  \label{Pt_v11}
\end{equation} 
which can be rewritten as
\begin{equation}
    \tilde{D} \, Z > \frac{u_2}{A \rho} + \frac{1}{A \rho k_2}, \label{Pt_v12}
\end{equation}
where $\tilde{D}$ is given in Table I.  
It can be observed that, when $\tilde{D}<0$ then (\ref{Pt_v11}) does not hold. On the other hand, when $\tilde{D}>0$, i.e., $u_t (1+u_1 k_1)(1+k_2 u_2) + k_2 u_1 u_2 < 1$, (\ref{Pt_v11}) holds when $Z>\frac{u_2}{\tilde{D} A \rho} + \frac{1}{\tilde{D} A \rho k_2}$. Hence, considering all the above, $P_{t,1}$ can be calculated as presented in Table \ref{Pt_1_2} given at the top of the next page with $N$ given in Table \ref{param_values}.

  \begin{table*}
		\vspace{-0.3cm} 
  \linespread{0.5}\selectfont
              	\centering
              	\caption{$P_{t,1}$, $P_{t,2}$, $P_{t,11}$, $P_{t,12}$, $P_{t,21}$, $P_{t,22}$ } \vspace{-0.2cm}
					\scalebox{0.9}{\begin{tabular}{|c|c|}
						\hline
						\textbf{Term} & \textbf{Expression} \\
						\hline
						\hline
						$P_{t,1}$ & $ \begin{aligned}& \int_{\frac{u_2}{\tilde{D} A \rho}+\frac{1}{\tilde{D} A \rho k_2}}^{\infty}\int_{D_z}^{z (\frac{\eta}{A k_2 u_t}+\frac{\eta}{A k_2})\frac{1}{\frac{B}{A k_2 u_1}+\frac{B k_1}{A k_2}}} \left( F_{|h_2|^2}\left(\frac{B \rho y - \eta \rho z u_1 - u_1}{A \rho k_2 u_1}\right) 
- F_{|h_2|^2}\left( \frac{B \rho y u_2 + \eta \rho z u_2  +  u_2  }{A \rho}\right) \right) f_{|h_{1}|^2}(y) f_{Z}(z) \, \mathrm{d}y \, \mathrm{d}z
\\
&= - \underbrace{\int_{\frac{u_2}{\tilde{D} A \rho}+\frac{1}{\tilde{D} A \rho k_2}}^{\infty}\int_{D_z}^{ N z } e^{-\frac{B \rho y - \eta \rho u_1 z - u_1 }{A \rho k_2 \lambda_2 u_1}} f_{|h_{1}|^2}(y) f_{Z}(z) \mathrm{d}y \mathrm{d}z}_{P_{t,11}}   +    \underbrace{\int_{\frac{u_2}{\tilde{D} A \rho}+\frac{1}{\tilde{D} A \rho k_2}}^{\infty}\int_{D_z}^{ N z } e^{-\frac{B \rho u_2 y + \eta \rho u_2 z + u_2 }{A  \lambda_2 \rho}} f_{|h_{1}|^2}(y) f_{Z}(z) \mathrm{d}y \mathrm{d}z}_{P_{t,12}} \end{aligned}$  \\
						\hline
						$P_{t,2}$ &  $\begin{aligned}&\int_{- \frac{1}{K} \left(\frac{u_2}{A \rho}+\frac{1}{A \rho k_2}\right)\frac{1}{\frac{B k_1}{A k_2}+\frac{B u_2}{A}}}^{\infty}\int_{N z}^{- \frac{z(\eta u_2-\frac{\eta}{ k_2 u_t})+\frac{u_2}{ \rho}+\frac{1}{ \rho k_2}}{\frac{B k_1}{ k_2}+ B u_2}} \left( - F_{|h_2|^2} \left(\frac{B \rho y u_2 + \eta \rho z u_2  +  u_2  }{A \rho}  \right)     + F_{|h_2|^2} \left( \frac{\eta \rho z - B k_1 \rho y u_t - u_t}{A \rho k_2 u_t}\right) \right) f_{|h_{1}|^2}(y) f_{Z}(z) \, \mathrm{d}y \, \mathrm{d}z
\\
& \ \ \ \ \ \ \ \ \ \ \ \ \ \ \ \ \ \ \ \ \ \ \ \ =  \underbrace{\int_{- \frac{1}{K} \left(\frac{u_2}{A \rho}+\frac{1}{A \rho k_2}\right)\frac{1}{\frac{B k_1}{A k_2}+\frac{B u_2}{A}}}^{\infty}   \int_{N z}^{- \frac{z(\eta u_2-\frac{\eta}{ k_2 u_t})+\frac{u_2}{ \rho}+\frac{1}{ \rho k_2}}{\frac{B k_1}{ k_2}+ B u_2}}  e^{-\frac{B \rho u_2 y + \eta \rho u_2 z + u_2}{A \lambda_2 \rho}} f_{Z}(z) f_{|h_{1}|^2}(y) \mathrm{d}y \mathrm{d}z \mathrm{d}y}_{P_{t,21}}
\\ 
& \ \ \ \ \ \ \ \ \ \ \ \ \ \ \ \ \ \ \ \ \ \ \ \ -    \underbrace{\int_{- \frac{1}{K} \left(\frac{u_2}{A \rho}+\frac{1}{A \rho k_2}\right)\frac{1}{\frac{B k_1}{A k_2}+\frac{B u_2}{A}}}^{\infty}   \int_{N z}^{- \frac{z(\eta u_2-\frac{\eta}{ k_2 u_t})+\frac{u_2}{ \rho}+\frac{1}{ \rho k_2}}{\frac{B k_1}{ k_2}+ B u_2}}  e^{-\frac{\eta \rho z - B k_1 \rho y u_t - u_t}{A \rho k_2 \lambda_2 u_t}} f_{Z}(z) f_{|h_{1}|^2}(y) \mathrm{d}y \mathrm{d}z \mathrm{d}y}_{P_{t,22}}\end{aligned}$  \\
						\hline
						$P_{t,11}$ & $ - \frac{ A k_2 \lambda_2 u_1 e^{\frac{ 1}{A \rho k_2 \lambda_2 }} }{ A k_2 \lambda_2 u_1 + B \lambda_1 }   \left(  \int_{\frac{u_2}{\tilde{D} A \rho}+\frac{1}{\tilde{D} A \rho k_2}}^{\infty} e^{- Q_3 z } f_{Z}(z) \mathrm{d}z     - e^{- (\frac{1}{\lambda_1}+\frac{B}{A k_2 \lambda_2 u_1}) (\frac{u_2}{A C \rho}+\frac{1}{A C \rho k_2})   }  \int_{\frac{u_2}{\tilde{D} A \rho}+\frac{1}{\tilde{D} A \rho k_2}}^{\infty} e^{- Q_4 z  } f_{Z}(z) \mathrm{d}z    \right)$  \\
						\hline
						$P_{t,12}$ & $ - \frac{ A \lambda_2  e^{-\frac{ u_2}{A  \lambda_2 \rho }} }{ A \lambda_2  + B u_2 \lambda_1 } \left(    \int_{\frac{u_2}{\tilde{D} A \rho}+\frac{1}{\tilde{D} A \rho k_2}}^{\infty} e^{-Q_5 z} f_{Z}(z) \mathrm{d}z   - e^{- (\frac{1}{\lambda_1}+\frac{B u_2}{A \lambda_2 }) (\frac{u_2}{A C \rho}+\frac{1}{A C \rho k_2})   }  
  \int_{\frac{u_2}{\tilde{D} A \rho}+\frac{1}{\tilde{D} A \rho k_2}}^{\infty}  e^{- Q_6 z   }   f_{Z}(z) \mathrm{d}z    \right)$   \\
						\hline
						$P_{t,21}$ & $ - \frac{A \lambda_2  e^{-\frac{u_2}{A \lambda_2 \rho}}}{A \lambda_2  + B \lambda_1 u_2}  \left(  e^{ (\frac{1}{\lambda_1}+\frac{B u_2}{A \lambda_2 })(\frac{u_2}{\rho}+\frac{1}{\rho k_2})\frac{1}{\frac{B k_1}{k_2}+B u_2}} \int_{- \frac{1}{K} \left(\frac{u_2}{A \rho}+\frac{1}{A \rho k_2}\right)\frac{1}{\frac{B k_1}{A k_2}+\frac{B u_2}{A}}}^{\infty} e^{- Q_7 z }     f_{Z}(z) \mathrm{d}z        -  \int_{- \frac{1}{K} \left(\frac{u_2}{A \rho}+\frac{1}{A \rho k_2}\right)\frac{1}{\frac{B k_1}{A k_2}+\frac{B u_2}{A}}}^{\infty} e^{- Q_{5} z }     f_{Z}(z) \mathrm{d}z  \right) $  \\
						\hline
                   $P_{t,22}$ & $- \frac{A k_2 \lambda_2  e^{\frac{ 1}{A \rho k_2 \lambda_2}}}{A k_2 \lambda_2  - B k_1 \lambda_1}   \left(  e^{ (\frac{1}{\lambda_1}-\frac{B k_1}{A k_2 \lambda_2})(\frac{u_2}{\rho}+\frac{1}{\rho k_2})\frac{1}{\frac{B k_1}{k_2}+B u_2}}  \int_{- \frac{1}{K} \left(\frac{u_2}{A \rho}+\frac{1}{A \rho k_2}\right)\frac{1}{\frac{B k_1}{A k_2}+\frac{B u_2}{A}}}^{\infty}  e^{- Q_{8} z } f_{Z}(z) \mathrm{d}z  -  \int_{- \frac{1}{K} \left(\frac{u_2}{A \rho}+\frac{1}{A \rho k_2}\right)\frac{1}{\frac{B k_1}{A k_2}+\frac{B u_2}{A}}}^{\infty}  e^{- Q_{9} z} f_{Z}(z) \mathrm{d}z \right) $   \\ 
                   \hline
				\end{tabular} }		\label{Pt_1_2}
				\hfill
				\vspace{-0.7cm}
      \end{table*}

Regarding $P_{t,2}$, it is obvious that for (\ref{Pt_v8}) to take non-zero values, the following must hold
\begin{equation}
    \begin{split}
      \frac{B \rho |h_1|^2 u_2 + \eta \rho Z u_2  +  u_2  }{A \rho} < \frac{\eta \rho Z - B k_1 \rho |h_1|^2 u_t - u_t}{A \rho k_2 u_t},
    \end{split}  \label{Pt_v13}  
\end{equation}
which can be equivalently written as
\begin{equation}
|h_1|^2 < - \frac{Z(\eta u_2-\frac{\eta}{ k_2 u_t})+\frac{u_2}{ \rho}+\frac{1}{ \rho k_2}}{\frac{B k_1}{ k_2}+ B u_2}.  \label{Pt_v14}
\end{equation}
Taking into account the condition of (\ref{Pt_v14}) as well as the first term of the probability in (\ref{Pt_v8}), we must determine for which $Z$ values, the following inequality holds
\begin{equation}
    \frac{Z (\frac{\eta}{A k_2 u_t}+\frac{\eta}{A k_2})}{\frac{B}{A k_2 u_1}+\frac{B k_1}{A k_2}} <  - \frac{Z(\eta u_2-\frac{\eta}{ k_2 u_t})+\frac{u_2}{ \rho}+\frac{1}{ \rho k_2}}{\frac{B k_1}{ k_2}+ B u_2},
\end{equation}
or equivalently for which $Z$ values the below inequality holds 
\begin{equation}
  K Z < - \left(\frac{u_2}{ \rho}+\frac{1}{ \rho k_2}\right)\frac{1}{\frac{B k_1}{ k_2}+B u_2}, \label{Pt_v16}
\end{equation}
where $K$ 
is given in Table I. It is obvious that when $K\!>\!0$, there are no $Z$ values that allow (\ref{Pt_v16}) to hold. On the contrary, when $K\!<\!0$, i.e., $k_2 u_2 u_1 ( 1  \!+\! u_t \!+\! k_1 u_t ) \!+\! u_t (k_1 u_1 \!+ \!k_2 u_2) \!<\! 1$, then (\ref{Pt_v16}) holds when $   Z > - \frac{1}{K} \left(\frac{u_2}{ \rho}+\frac{1}{ \rho k_2}\right)\frac{1}{\frac{B k_1}{ k_2}+B u_2}=\frac{u_2}{\tilde{D} A \rho} + \frac{1}{\tilde{D} A \rho k_2}$. Considering all the above, $P_{t,2}$ can be calculated as presented in Table \ref{Pt_1_2}. 

At this point, it is highlighted that by substituting $f_{|h_{1}|^2}(y)$ in $P_{t,11}$, $P_{t,12}$, $P_{t,21}$ and $P_{t,22}$ that appear in the first two rows of Table \ref{Pt_1_2} and after some algebraic manipulations, $P_{t,11}$, $P_{t,12}$, $P_{t,21}$ and $P_{t,22}$ can be transformed as shown in Table III. By rewriting them via function $\Phi(\alpha, \beta)$ presented in Lemma \ref{lem:G}, then invoking
$P_{t,11}$, $P_{t,12}$ in $P_{t,1}$ and $P_{t,21}$, $P_{t,22}$ in $P_{t,2}$ and then the resultants $P_{t,1}$, $P_{t,2}$ in (\ref{Pt_v6}) and finally the resultant (\ref{Pt_v6}) in (\ref{OP_xt_v1}), we get (\ref{xt_OP}). This concludes the proof.

\vspace{-0.2cm}
\section*{Appendix C: Proof of theorem \ref{thm:IP_BD}}
\vspace{-0.1cm}
Leveraging the law of total probability, BD's IP becomes
\begin{equation}
	\begin{split}
		P^{\text{int}}_{t}&=\sum_{k=1}^{2} \Pr\left(\underset{i \in \{1,2\}}{\text{rand}}{U_i}=U_k\right) \\
		&\times  \Pr\left(\underset{j \in \{1,...,M\}}{\text{max}}\left\{\frac{ \eta \rho |h_{tj}|^2 (|h_{2t}|^2+|h_{1t}|^2)}{ a_2 \rho|h_{kj}|^2  + 1}\right\}>u_t^{\text{int}}\right) \\
  & =\frac{1}{2} \sum_{k=1}^{2}  \Pr\left(\underset{j \in \{1,...,M\}}{\text{max}}\left\{V_{kj}\right\}>\frac{u_t^{\text{int}}}{(|h_{2t}|^2+|h_{1t}|^2)}\right) \\
  & =\frac{1}{2} \sum_{k=1}^{2}\! \left(\! 1 \!-\!  \Pr\!\left(\!\underset{j \in \{1,...,M\}}{\text{max}}\left\{V_{kj}\right\}<\frac{u_t^{\text{int}}}{(|h_{2t}|^2+|h_{1t}|^2)}\!\right) \! \right) \!,
	\end{split}   \label{IP_Ut_v0}
\end{equation}
where 
\begin{equation}
	V_{kj}= \frac{\eta \rho |h_{tj}|^2}{ a_2 \rho|h_{kj}|^2  + 1}, \ \ \ k \in \{1,2\}.
\end{equation}
The CDF of $V_{kj}$ can be calculated in a similar manner to the CDF of the RV $Y^{I}_{j}$ provided in (\ref{F_Y1}), thus 
\begin{equation}
	\begin{split}
		F_{V_{kj}}(v)= 1 - \frac{\eta \lambda_{tj} e^{-\frac{v}{\eta \rho \lambda_{tj}}}}{\eta \lambda_{tj} + a_2 \lambda_{kj} v}. 
	\end{split} \label{F_V}
\end{equation}
Leveraging (\ref{F_V}) and (\ref{CDF_MAX}), (\ref{IP_Ut_v0}) can be transformed into 
\begin{equation}
	\begin{split}
		P^{\text{int}}_{t}&= 1 - \frac{1}{2}  \sum\limits_{k=1}^{2}  \underbrace{\int_{0}^{\infty} \prod_{j=1}^{M} F_{V_{kj}}\!\left(\frac{u_t^{\text{int}}}{w}\right) f_W(w) \, \mathrm{d}w}_{I_{kj}},
	\end{split}   \label{IP_Ut}
\end{equation}
where $f_W(w)$ has been provided in (\ref{f_W}). 

Next, we focus on the calculation of $I_{kj}$. When $\lambda_{1t} \ne \lambda_{2t}$, by substituting the first branch of $f_W(w)$, it occurs
\begin{equation}
\begin{split}
 I_{kj} &\!=\! \frac{1}{\lambda_{1t}\!-\!\lambda_{2t}}\! \sum\limits_{i=1}^{2} (-1)^{i+1} \!\int_{0}^{\infty} \! e^{-\frac{w}{\lambda_{it}}} \!  \prod_{j=1}^{M}\! F_{V_{kj}}\!\left(\!\frac{u_t^{\text{int}}}{w}\!\right) \!  \mathrm{d}w. 
 \label{I_kj_1} 
 \end{split} 
\end{equation}
Substituting (\ref{F_V}) into (\ref{I_kj_1}), it occurs that the resultant integral cannot be solved in closed-form. Hence, the well-known Gauss-Laguerre quadrature \cite{hildebrand1987introduction} is exploited for its approximation. This way, the first branch of (\ref{Ikj_f}) occurs.

On the contrary, when $\lambda_{1t}=\lambda_{2t}$, by applying the second branch of $f_W(w)$ into (\ref{IP_Ut}), $I_{kj}$ can be rewritten as
\begin{equation}
\begin{split}
 I_{kj} &= \frac{1}{\lambda^2} \int_{0}^{\infty}  w \, e^{-\frac{w}{\lambda}}  \prod_{j=1}^{M} F_{V_{kj}}\!\left(\frac{u_t^{\text{int}}}{w}\right)  \mathrm{d}w.  
 \label{I_kj_v2} 
 \end{split}
\end{equation}
Exploiting the Gauss-Laguerre quadrature for estimating (\ref{I_kj_v2}), we get the second branch of (\ref{Ikj_f}). By invoking (\ref{Ikj_f}) in (\ref{IP_Ut}), (\ref{IP_Ut_vv0}) is derived. This completes the proof.

\vspace{-0.5cm}

\bibliographystyle{IEEEtran} 
\bibliography{IEEEabrv,References2}
\balance

\end{document}